\documentclass[sigconf]{acmart}
\AtBeginDocument{%
  }

\copyrightyear{2025}
\acmYear{2025}
\setcopyright{cc}
\setcctype{by}
\acmConference[IUI '25]{30th International Conference on Intelligent User Interfaces}{March 24--27, 2025}{Cagliari, Italy}
\acmBooktitle{30th International Conference on Intelligent User Interfaces (IUI '25), March 24--27, 2025, Cagliari, Italy}
\acmDOI{10.1145/3708359.3712147}
\acmISBN{979-8-4007-1306-4/25/03}

\usepackage{tabu}                      % only used for the table example
\usepackage{csquotes}
\usepackage{booktabs}                  % only used for the table example
\usepackage{lipsum}                    % used to generate placeholder text
\usepackage{mwe}                       % used to generate placeholder figures
\usepackage{graphicx}
\usepackage{color}
\usepackage{balance}
\usepackage{setspace}

\newcommand{\revise}[1]{#1}

\begin{document}


\title{\systemname: A Visual Analytics Workflow for Subgroup-based Semantic Error Analysis of CVML Models}

\author{Jun Yuan}
\orcid{0000-0003-1952-5221}
\affiliation{%
  \institution{Apple}
  \city{Cupertino}
  \state{CA}
  \country{USA}
}
\email{junyuan@apple.com}

\author{Kevin Miao}
\orcid{0000-0001-9151-3028}
\affiliation{%
  \institution{Apple}
   \city{Cupertino}
  \state{CA}
  \country{USA}
}
\email{kevin_miao@apple.com}

\author{Heyin Oh}
\orcid{0009-0000-3523-637X}
\affiliation{%
  \institution{Apple}
   \city{Cupertino}
  \state{CA}
  \country{USA}
}
\email{heyin_oh@apple.com}

\author{Isaac Walker}
\orcid{0009-0002-1411-2241}
\affiliation{%
  \institution{Apple}
   \city{Cupertino}
  \state{CA}
  \country{USA}
}
\email{isaac_walker@apple.com}

\author{Zhouyang Xue}
\orcid{0009-0009-5735-038X}
\affiliation{%
  \institution{Apple}
   \city{Cupertino}
  \state{CA}
  \country{USA}
}
\email{zhouyang_xue@apple.com}

\author{Tigran Katolikyan}
\orcid{0009-0003-6276-9659}
\affiliation{%
  \institution{Apple}
   \city{Cupertino}
  \state{CA}
  \country{USA}
}
\email{tigran_katolikyan@apple.com}

\author{Marco Cavallo}
\orcid{0000-0003-1506-4536}
\affiliation{%
  \institution{Apple}
   \city{Cupertino}
  \state{CA}
  \country{USA}
}
\email{marco_cavallo@apple.com}

\renewcommand{\shortauthors}{ Yuan et al.}

\begin{abstract}
   % @Jun: I just realized we haven't updated the abstract yet :P
Effective error analysis is critical for the successful development and deployment of CVML models. One approach to understanding model errors is to summarize the common characteristics of error samples. This can be particularly challenging in tasks that utilize unstructured, complex data such as images, where patterns are not always obvious. Another method is to analyze error distributions across pre-defined categories, which requires analysts to hypothesize about potential error causes in advance. Forming such hypotheses without access to explicit labels or annotations makes it difficult to isolate meaningful subgroups or patterns, however, as analysts must rely on manual inspection, prior expertise, or intuition. This lack of structured guidance can hinder a comprehensive understanding of where models fail. 
To address these challenges, we introduce VibE, a semantic error analysis workflow designed to identify where and why computer vision and machine learning (CVML) models fail at the subgroup level, even when labels or annotations are unavailable. VibE incorporates several core features to enhance error analysis: semantic subgroup generation, semantic summarization, candidate issue proposals, semantic concept search, and interactive subgroup analysis. By leveraging large foundation models (such as CLIP and GPT-4) alongside visual analytics, VibE enables developers to semantically interpret and analyze CVML model errors. This interactive workflow helps identify errors through subgroup discovery, supports hypothesis generation with auto-generated subgroup summaries and suggested issues, and allows hypothesis validation through semantic concept search and comparative analysis. Through three diverse CVML tasks and in-depth expert interviews, we demonstrate how VibE can assist error understanding and analysis.

\end{abstract}




\begin{teaserfigure}
  \includegraphics[width=\textwidth]{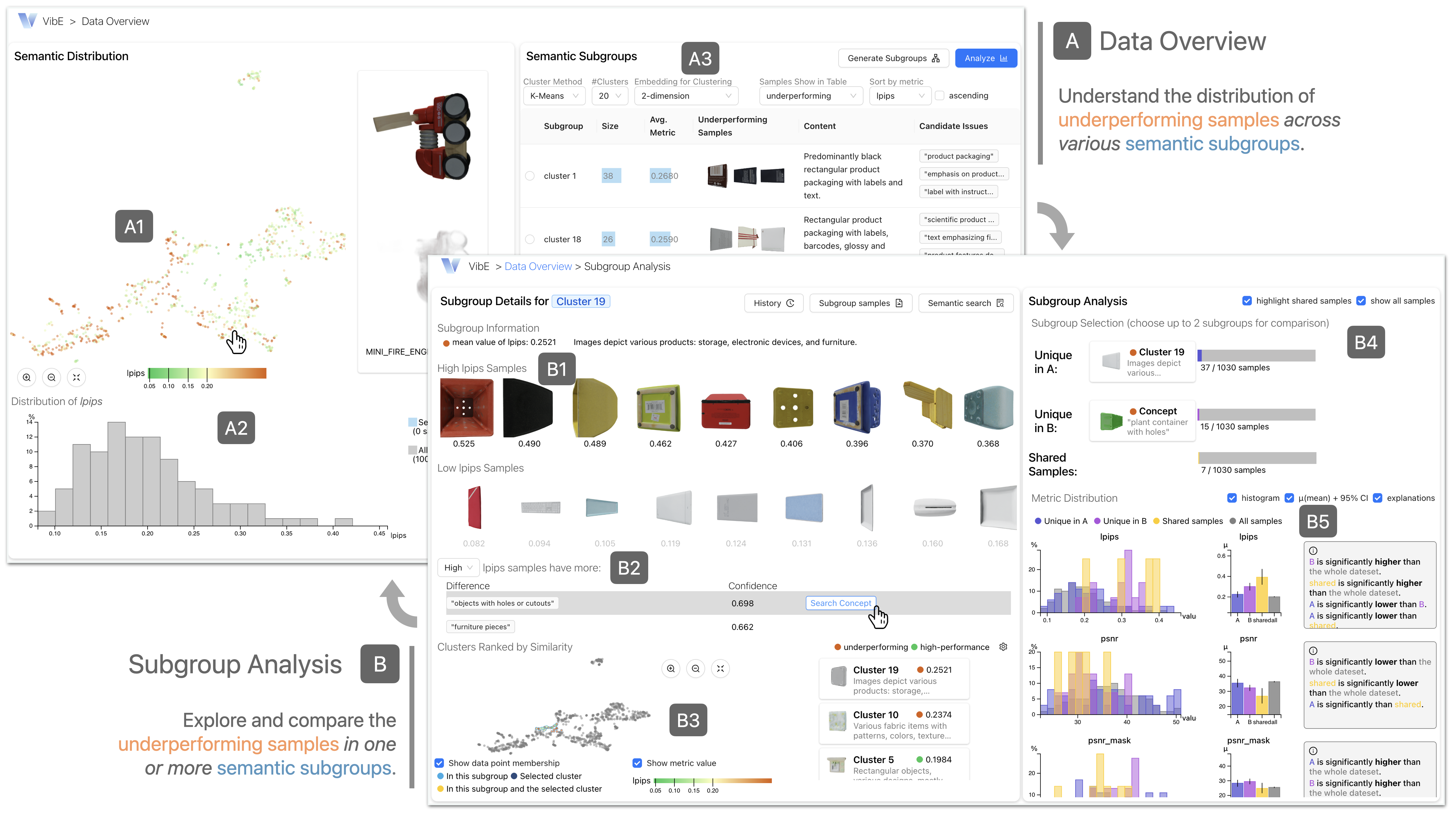}
  \caption{{\systemname} enables CVML model developers to analyze the semantic distribution of model errors with the support of large foundation models. It offers two main pages: (A) the \textit{Data Overview} page, where users can explore error samples and review system-generated clusters of semantically similar samples; and (B) the \textit{Subgroup Analysis} page, which helps users generate and validate hypotheses about error-related semantic features.}
  \Description{The user interface of \systemname{} to assist semantic error analysis of a 3d asset generation model.}
  \label{fig:ui_3d}
\end{teaserfigure}


\newcommand{\systemname}{VibE}
\newcommand*{\gear}{\includegraphics[width=10pt,height=10pt]{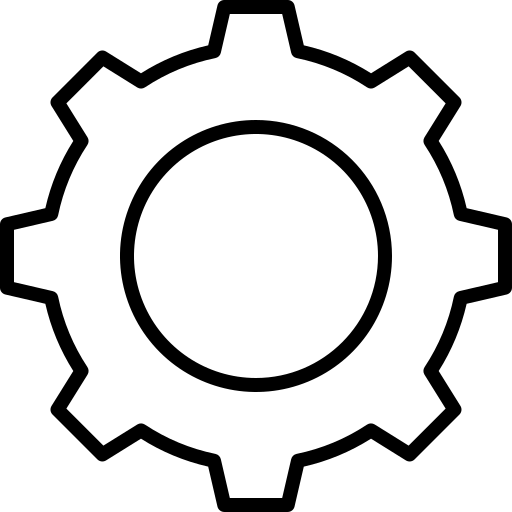}}%
\newcommand{\colorbtn}{[\includegraphics[width=10pt, height=10pt]{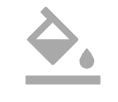}]}
\newcommand{\filterbtn}{[\includegraphics[width=10pt, height=10pt]{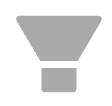}]}
\newcommand{\lookupbtn}{[\includegraphics[width=10pt, height=10pt]{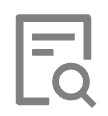}]}

\definecolor{theblue}{RGB}{152,181,211}
\definecolor{theorange}{RGB}{255, 124, 54}

\definecolor{bluetextcolor}{RGB}{99, 136, 181}
\definecolor{yellowtextcolor}{RGB}{166, 126, 58}
\definecolor{greentextcolor}{RGB}{107, 158, 113}

\newcommand{\fixmepls}[1]{\textbf{\textcolor{red}{#1}}}
\newcommand{\textorange}[1]{\textcolor{theorange}{#1}}
\newcommand{\textblue}[1]{\textcolor{theblue}{#1}}

\newcommand{\dataflowblue}[1]{\textcolor{bluetextcolor}{#1}}
\newcommand{\dataflowgreen}[1]{\textcolor{greentextcolor}{#1}}
\newcommand{\dataflowyellow}[1]{\textcolor{yellowtextcolor}{#1}}

\maketitle
\section{Introduction}
In recent years, Computer Vision and Machine Learning (CVML) algorithms have advanced rapidly, transforming fields like healthcare, retail, and finance~\cite{szeliski2022computer}. As these models become more complex, ensuring their performance, reliability, and fairness at scale has increasingly become challenging. A critical aspect of evaluating these models is \textit{subgroup error analysis}, examining model performance across specific subgroups---distinct subsets of data. Each subgroup is defined by common characteristics within the group, like demographics (e.g., age, gender, ethnicity) or pre-defined labels (e.g., image classes). The goal is to uncover biases or disparities in model performance that may not be apparent at the model level. This analysis ultimately helps uncover issues, such as overfitting to certain patterns, poor generalizability, or unfair outcomes in minority groups, which are key factors in improving model robustness and fairness.

Traditional subgroup error analysis uses techniques such as confusion matrices and per-class accuracy~\cite{naidu2023review}. However, as we discuss in our formative study (Sec.~\ref{sec:workflow}), these methods depend heavily on the presence of complete metadata, clean labels, and well-defined taxonomies. In the absence of such data, these conventional techniques frequently fail to capture the nuanced factors driving performance variations. This is particularly true when samples contain anomalies or unknown characteristics, so-called ``aggressors''. For example, a dataset with demographic metadata might not necessarily contain features like tattoos or head coverings, which could have a big influence on the final model performance. This underscores the first major challenge: \textbf{identifying and analyzing meaningful subgroups} when traditional methods struggle with incomplete data or unexpected characteristics. 

Our study also highlights a second challenge: \textbf{validating error hypotheses}. Once a problematic subgroup is identified, determining whether the errors are specific to that subgroup or more widespread is difficult. 
Traditional workflows offer limited tools to explore correlations between subgroups and error patterns, making it hard to isolate the root causes of the model's shortcomings. \revise{While methods such as TCAV~\cite{pmlr-v80-kim18d} and self-supervised visual concept segments~\cite{he2022self} are able to assist users in exploring data samples with specific concepts, they depend on pre-generated concepts and, in some cases, require access to model internals, which may not always be available for error analysis.} 
% \revise{Although concept-based methods such as TCAV~\cite{pmlr-v80-kim18d}, which can generate high-dimensional representations of a "concept" learned by a model, are available, they require access to the model's internals and rely on human annotations to identify the "concepts" learned by the model. Furthermore, a "concept" or "aggressor" may not be directly learned or utilized by a model, but still needs to be analyzed. For example, an image relighting model may not perform well with "green apples" in an image, but this does not imply that the model has learned the concept of "green apples" in the rendering process. Instead, more complex underlying issues could be involved, which also requires an effective approach to validate such error cases.}

To address these challenges, we introduce a more \textit{semantic} approach to subgroup analysis, leveraging advances in self-supervised learning and large foundation models. \revise{Multimodal models} like CLIP~\cite{radford2021clip} \revise{and BLIP~\cite{li2022blip}} can extract rich, latent features from raw data, even without explicit labels. These models allow us to align visual and textual data, creating semantically meaningful subgroups, and enabling a more nuanced analysis of subgroup errors. \revise{We further leverage the state-of-the-art multimodal large language models (LLMs), such as GPT-4~\cite{GPT4}, to assist users to understand the problematic subgroups and candidate issues with these subgroups.}

\revise{More specifically, our} work builds on a formative study with seven CVML engineers (Sec.\ref{sec:study}) and a hierarchical task analysis of their debugging workflows (Fig.\ref{fig:hta}). From this, we derived design goals to develop semantic-focused methodologies for subgroup analysis (Sec.\ref{sec:method}). We introduce \systemname{}, an intelligent workflow powered by large foundation models that facilitates the identification of semantic subgroups, discovery of potential data issues, retrieval of relevant samples, and recommendation of similar subgroups for hypothesis validation. In Sec.\ref{sec:system}, we present the implementation of this workflow as a visual analytics system. This system supports developers in investigating \textit{where} their models make errors, forming hypotheses about \textit{why} these errors occur, and verifying \textit{whether} the identified factors are statistically likely to contribute to the mistakes.

We evaluate our approach through three case studies involving expert interviews with 3D asset generation, face reconstruction, and object detection models. Our findings show that foundation models enhance error analysis by providing meaningful data summaries and supporting semantic exploration. Additionally, visual analytics techniques help mitigate the limitations of foundation models, such as uncertainty or ambiguity in their outputs. We also discuss lessons learned and offer design guidelines for integrating foundation models into visual analytics systems while addressing our system’s limitations for real-world use.

In summary, our contributions are: (1) an initial study with seven CVML engineers, which informed key design principles; (2) a novel analytic workflow that supports semantic subgroup identification and error validation; (3) the development of a visual analytics system implementing this workflow; (4) three case studies, with expert interviews, each focusing on a different CVML task; and (5) insights gained from integrating foundation models into a visual analytics system.

\section{Background and Related Work}
\label{sec:related}
In this work, we propose a data-centric visual analytics workflow that empowers model developers to formulate and validate hypotheses, leading to insights into model behavior and performance discrepancies through the examination of semantically relevant data subgroups. Our research intersects with various related fields. In this section, we first review existing studies on model debugging tasks that have informed our preliminary interviews (Sec 2.1), providing a deeper understanding of the problem space. Next, we discuss the background and motivation for applying large foundation models to visual analytics tasks (Sec 2.2). Finally, we explore existing tools for model error analysis at the subgroup level (Sec 2.3), particularly those that assist users in semantically understanding model behaviors.

\subsection{Understanding Model Debugging Workflows}

Model interpretation and behavior analysis have garnered significant attention in the HCI and visualization communities, leading to the development of numerous tools designed to support various analytical tasks across different machine learning models~\cite{hohman2018visual,yuan2021survey}. Hohman et al.~\cite{hohman2018visual} explore state-of-the-art visual analytics techniques for deep learning models by addressing \textit{Five W's and How} questions. Yuan et al.~\cite{yuan2021survey} break down the machine learning pipeline into six stages, categorizing visual analytics systems based on these stages and the analytical goals they aim to achieve. Subramonyam and Hullman~\cite{subramonyam2023we} further classify visual analytics systems by considering factors such as expert involvement, prior knowledge, and specific tasks. In a more recent survey, La Rosa~\cite{la2023state} emphasizes visual analytics design for explainable AI (XAI) methods, categorizing techniques based on their support for different users and tasks. 

With the advancement of XAI techniques, numerous empirical studies have explored human-AI decision-making~\cite{lai2021towards} and conducted interviews with domain experts to examine industry practices, challenges, and needs~\cite{hong2020human,liao2020questioning}. However, these studies often concentrate on the general decision-making process~\cite{lai2021towards}, the entire model development pipeline~\cite{hong2020human}, or design concerns from UX and design practitioners~\cite{liao2020questioning, moore2023failurenotes}. Despite this, there remains a gap in systematic analyses that focus on the specific tasks and challenges involved in model understanding and debugging in practice. In this work, we address this gap by presenting a task analysis based on interviews with model developers to better understand their current debugging workflows and pain points.

\subsection{Usage of Foundation Models for VA}
Large language models like Gemini, GPT, and Llama are billion-parameter transformer models trained on petabytes of internet data~\cite{geminiteam2023gemini, openai2023gpt4, touvron2023llama}. Recent advancements in natural language processing have propelled these models to prominence, thanks to their remarkable generalization capabilities, zero-shot learning, and emergent properties. In turn, various use cases have emerged where LLMs assist in data exploration and analysis~\cite{ma2023demonstration, chen2023genspectrum}. More recently, multimodal models such as Llava~\cite{liu2023visual}, GPT-4~\cite{openai2023gpt4}, Flamingo~\cite{alayrac2022flamingo}, and Gemini~\cite{geminiteam2023gemini} have demonstrated even greater power by reasoning over both visual and textual inputs. Given their capabilities, visual analytics (VA) researchers have developed diverse applications using these foundation models. LEVA~\cite{zhao2024leva}, for example, leverages GPT-4~\cite{openai2023gpt4} to recommend insights in VA workflows, while vision-language models like CLIP~\cite{radford2021clip} are employed to help users interpret the semantic meaning of image data~\cite{hoque2022visual, jamonnak2023ow}. Yang et al.~\cite{yang2023foundation} further explored how VA techniques support the development of foundation models and how these models can, in turn, enhance VA tasks. In our work, we utilize foundation models to assist in model debugging by leveraging semantically meaningful features, such as CLIP embeddings, for subgroup identification and automatic summarization. We also reflect on how participants in our case studies (Sec.~\ref{sec:study}) employed insights from foundation models during VA tasks, and we collect feedback from domain experts regarding their perspectives on the utility and limitations of these models.

\subsection{Semantic Error Analysis at the Subgroup Level}
Understanding data patterns at the subgroup level is important for visual analytics, as user-defined areas of interest frequently exhibit clear semantic meaning ~\cite{shrinivasan2010supporting}. Consequently, many VA systems are designed to facilitate data analysis over user-defined subsets of interest. For instance, TaxiVis ~\cite{ferreira2013visual} supports filtering and brushing operations, enabling users to analyze spatio-temporal data patterns within specific queried subsets. Below, we outline several research efforts focused on subgroup analysis for text and image data, respectively.

% current work on NLP, tabular data... and how they are relevant to semantics
\subsubsection{Semantic error analysis for NLP and tabular data.} 
When analyzing model behavior and errors, generating groups of data samples with shared characteristics can provide valuable insights. These subgroups, also known as ``slices''~\cite{chung2019slice}, ``context'' ~\cite{yuan2022context}, or ``neighborhood''~\cite{molnar2022}, help in understanding patterns within the data. In the visual analytics (VA) community, subgroups can be created in various ways. For instance, SliceFinder uses decision tree rules to form similar subgroups ~\cite{chung2019slice}, while FairVis ~\cite{cabrera2019fairvis} helps users generate and interpret subgroups by clustering and highlighting ``important'' features. Zeno ~\cite{cabrera2023zeno} offers Python APIs for building these subgroups using modular components.

Other systems focus on generating subgroups based on neuron activations ~\cite{lee2022viscuit}, or local model explanations ~\cite{yuan2022subplex}. In natural language processing (NLP), CheckList ~\cite{ribeiro2020beyond} generates new subsets of documents to assess model robustness, and Errudite ~\cite{wu2019errudite} enables querying subgroups of documents based on token and metadata attributes. iSEA ~\cite{yuan2022isea} creates token-based rules to identify error-prone subgroups for further analysis.

For tabular data, features often have intrinsic semantic meanings (e.g., gender, time, count). Similarly, in text data, tokens (words) and metadata (e.g., document length, question type) provide semantic value. To understand the meaning behind subgroups in tabular or text data, visualizations often emphasize the distribution of meaningful features. Tools like parallel coordinates assist in exploring these data subsets ~\cite{andrienko2004parallel}. More recent VA research has used if-then rules in combination with model statistics (e.g., data size, error rates) to describe consistent model behavior within specific data subsets ~\cite{ming2018rulematrix, yuan2022isea}.

\subsubsection{Semantic error analysis for image data.} 
Extracting semantically meaningful subgroups for image data without human annotations remains a significant challenge. In some CVML tasks, class information is available through ground truth and model predictions (e.g., semantic segmentation). Visual analytics (VA) systems designed for CVML models often leverage these classes or metadata to generate subgroups (e.g., SliceTeller ~\cite{9906903}, \cite{chen2023unified}), or they explore the hierarchical structure of classes\cite{bilal2017convolutional}. In recent years, researchers have introduced methods like TCAV ~\cite{pmlr-v80-kim18d} to extract ``concepts'' learned by CVML models. For example, \revise{TCAV can represent a concept ``striped'' based on user collected images of zebra and then tests how such concept influences the output of a given neural network.} ConceptExtract ~\cite{zhao2021humanintheloop} expands on this idea by \revise{enabling users to explore such concepts through an interactive user interface.} 
% though it requires access to model internals and human input to identify positive/negative samples tied to model behavior. 
% TCAV itself also relies on human annotations for these concepts.
\revise{Compared with these approaches, \systemname{} is model-agnostic, does not rely on annotations.}
\revise{More recently, researchers proposed a self-supervised model to learn different concepts as image segments ~\cite{he2022self} from a dataset. Based on this technique, Hoque et al.~\cite{hoque2022visual} construct rules of visual concepts to explain model behaviors, Zhang et al.~\cite{zhang2024slicing} use pre-generated concepts to prompt ChatGPT to generate different descriptions for data retrieval. In our work, instead of using pre-generated concepts, we provide intelligent assistance to help users generate hypotheses through content summarization and candidate data issue proposals, and support hypothesis validation through natural language.}

With advances in vision-language models, VA researchers have increasingly turned to CLIP ~\cite{radford2021clip} to derive the semantic meaning of image segments that may influence model predictions. For instance, Hoque et al. ~\cite{hoque2022visual} use CLIP to label image segments crucial to model predictions, while OW-Adapter ~\cite{jamonnak2023ow} uses CLIP embeddings to help users label unknown classes for open-world object detection models. Our work extends these approaches by enabling its application to any CVML model without requiring access to its internals. Instead of precomputing concepts assumed to be relevant to model behavior, we allow users to dynamically generate hypotheses about which concepts are linked to model errors. This is achieved using GPT-4 ~\cite{GPT4} for data summarization and CLIP for generating semantically meaningful embeddings, helping users test whether a specific concept is associated with model performance issues.

\section{Formative Study and Design Goals}
\label{sec:workflow}

We designed and conducted formative interviews with seven machine learning engineers (MLEs) to gain a deeper understanding of current error analysis practices to inform the design of our workflow. Based on these formative interviews, we performed a hierarchical task analysis~\cite{preece1994human} and task abstraction~\cite{lam2017bridging} focusing on the error analysis workflow. This analysis revealed two key pain points within the existing process. Based on our findings, we establish a set of design goals to address these challenges.

\subsection{Interview Procedure and Hierarchical Task Analysis}

\begin{figure*}
    \centering
    \includegraphics[width=\textwidth]{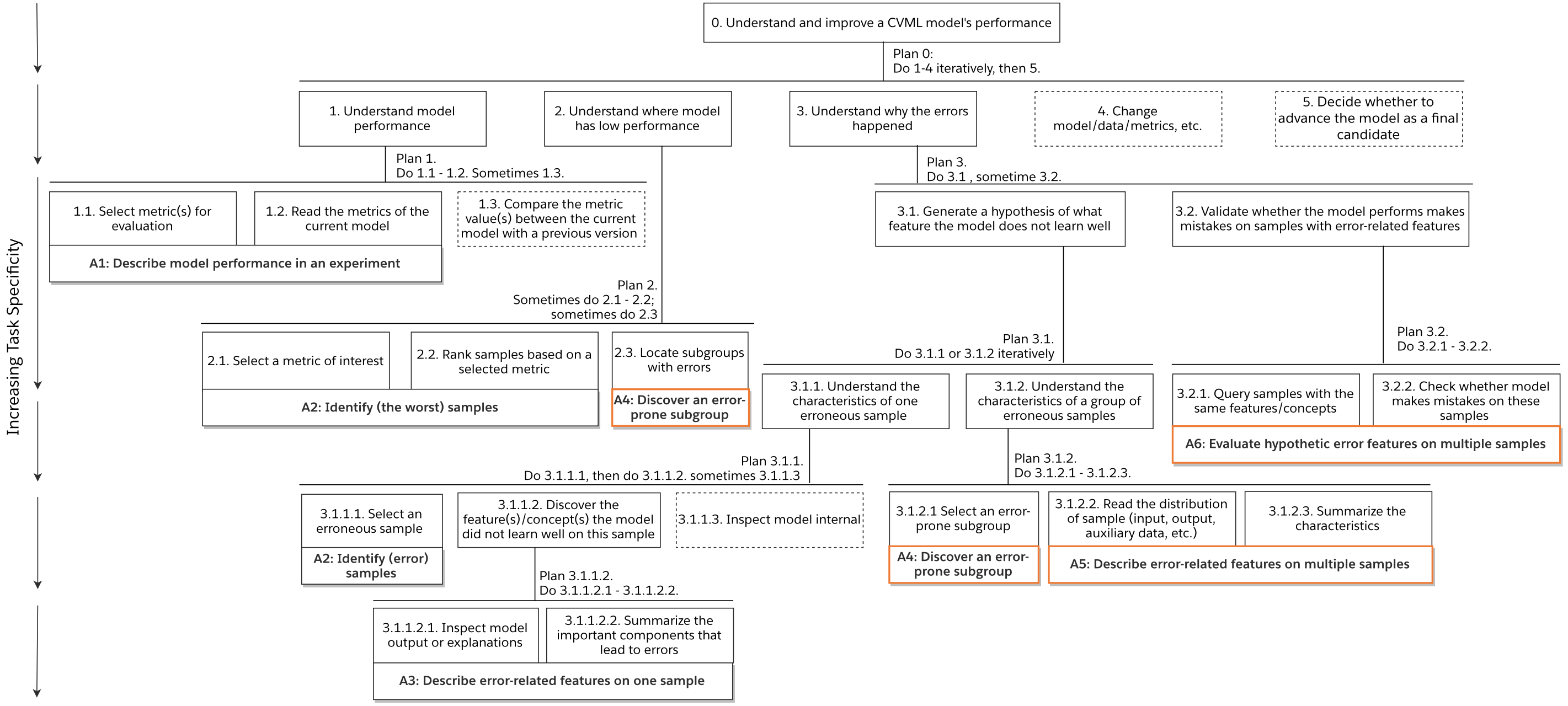}
    \caption{Hierarchical Task Abstraction (HTA) of model error analysis in order to improve model performance using box-and-line notation. We follow the standard conventions for hierarchical task analysis~\cite{kurniawan2004interaction} where tasks are represented by named boxes with a unique ID, which also indicates the hierarchical level of the task. Task abstraction based on ~\cite{lam2017bridging} are highlighted as A1-A6. The boxes with dashed lines are out of the scope of this paper. The \textorange{orange} boxes denote the abstract tasks that are relevant to pain points of conducting error analysis.}
    \label{fig:hta}
\end{figure*}

We conducted formative interviews with seven machine learning model developers (P1-7; 4 male, 3 female) from a tech company. Participants, aged 28 to 47, specialize in computer vision tasks involving image data as part of their daily responsibilities and possess 4 to 11 years of experience in the field. Their areas of expertise include face reconstruction (P1-3), semantic segmentation (P4-5), and object detection (P6-7). Each interview lasted between 30 and 45 minutes, with six conducted in person and one remotely.

The interviews were semi-structured, covering a range of topics, including the steps involved in model error analysis, the tools utilized, examples of errors, their causes, and proposed solutions. Based on the interview notes, we iteratively refined and validated our task analysis. Fig.~\ref{fig:hta} illustrates the resulting hierarchical task analysis. We have omitted the details of tasks 1.3 and 3.1.1.3 to focus on the model’s performance analysis of single test experiments, specifically from the perspective of the semantic meanings of the sample subgroups without accessing the model's internals, i.e. subgroup error analysis. We also exclude Tasks 4 and 5, which pertain to solutions and decisions made after the error analysis.

When analyzing model errors, an MLE typically initially identifies the potential causes of errors and subsequently, mitigate the issue to enhance the model performance. The process begins with an overview of the model's performance by reviewing key metrics that are significant to them (\textit{Task 1}). Next, they investigate which samples the model is struggling with (\textit{Task 2}) and explore the potential reasons underlying these mistakes (\textit{Task 3}). Underperforming samples may be ranked based on specific metrics (\textit{Tasks 2.1-2.2}) or grouped together as subgroups (e.g., samples from the same predefined class or other common characteristics; \textit{Task 2.3}). To understand the reasons behind sub-optimal performance, an MLE may either analyze a single error sample in detail (\textit{Task 3.1.1}) or summarize a group of underperforming samples (\textit{Task 3.1.2}). From this analysis, they formulate plausible hypotheses regarding the causes of the errors. We have found that MLEs seldom validate these hypotheses further (\textit{Task 3.2}); they often skip this step and make direct adjustments to the model (e.g., modifying model structure or parameters), data (e.g., collecting more data or refining annotations), or metrics (e.g., creating new metrics to better capture model behavior) (\textit{Task 4}). If they believe the adopted changes yield an adequate improvement, they will make decisions in incorporating the changes within the model in deployment (\textit{Task 5}).

\subsection{Outstanding Challenges in Semantic Error Analysis}
Our preliminary interviews emphasize the pivotal role of data in the model development process. Model errors are frequently attributed to data scarcity, a lack of diversity, and inconsistencies, which inhibit the model's ability to learn certain concepts and generalize effectively. Unlike other types of analyses where engineers may simply adjust the model architecture or tweak hyperparameters, the actionable outcomes often necessitate more extensive cross-functional efforts. In the case of in-house datasets, semantic analysis may lead to requests for additional data related to specific demographic segments or for annotations of samples with hypothesized aggressors affecting model performance. These efforts typically require significant time and resources, and their effectiveness is not guaranteed. Therefore, it is crucial to accurately assess which data issues should be prioritized for resolution. Based on our formative interviews and task analysis, we identify two major pain points in data-centric error analysis workflows.

\textbf{Difficulty Identifying Common Characteristics in underperforming Samples.} All participants reported using either a single underperforming sample (\textit{Abstract Task A3}) or a set of underperforming samples (\textit{Abstract Task A5}) to identify error-related features. Some participants (P4-7) work with supervised use cases that involve pre-defined labels corresponding to the content of a sample, while others (P1-3) do not. In instances where class information is available (e.g., object detection), samples often contain complex backgrounds, making it challenging to summarize their commonalities. For example, P7 described an analysis of a food detection model, where they first ranked the samples based on mean Intersection over Union (IoU) by object class. They focused on a sample labeled \textit{burger} that had a lower overall IoU score. When evaluating all burger samples with low IOU scores, they all displayed various other dishes on the background, including bread and side dishes. It remained unclear whether the issues stem from the fries, the bread, both or neither of them. To better assist MLEs in identifying error-related features, it is essential to improve the identification of subgroups containing semantically similar samples (\textit{Abstract Task A4}) and to provide support for subgroup summarization (\textit{Abstract Task A5}).

\textbf{Difficulty Generalizing Hypothetical Error Causes to Similar Samples Without Annotations.} Once a model developer identifies a hypothetical error cause, they typically make adjustments to their model, data, or metrics and then test whether these changes lead to improved performance. Often, this improvement necessitates requesting additional data. Before making such requests, model developers select samples with characteristics they suspect the model struggles with to validate their needs. For instance, P4 noted the challenge of manually selecting indoor images featuring specific layouts and environmental features not listed in the class labels. Similarly, P7 reported spending a significant amount of time handpicking images of objects not included in the predefined list, and cross-checking if they shared similar performance.
This process becomes even more difficult for tasks lacking class labels. P3 described spending $1-2$ hours handpicking images of interest and maintaining a JSON file containing the IDs of underperforming images with specific facial features. This labor-intensive workflow highlights the value model developers see in retrieving samples based on user-defined characteristics at scale during the error analysis process (\textit{Abstract Task A6}).

We particularly observe that the lack of adequate tooling for semantic-level performance analysis often led MLEs to either 1) tweak models and their input data based on a small number of problematic samples without sufficient verification, or 2) spend significant amounts of time manually gathering underperforming samples with shared characteristics. This effort was typically necessary to justify large requests to other teams responsible for data collection and annotation.

\subsection{Design Goals}
Building on the task analysis and key pain points identified, we propose the following design goals to guide the development of visual analytics systems aimed at assisting model developers in identifying and validating hypothetical error causes for CVML models.

\begin{itemize}

\item[G1.] \textbf{Facilitate the identification of semantic subgroups.} Analyzing underperforming samples all together can make it difficult to pinpoint potential error causes, as these data points might not share clear patterns. To address this, we propose complementing global inspection with the analysis of smaller data slices (subgroups) that share semantic characteristics. While users should have the flexibility to define their own subgroups, the system should also semi-automatically generate subgroups for easier exploration.

\item[G2.] \textbf{Provide a concise overview of each identified subgroup.} Understanding the characteristics, behaviors, and trends of a subgroup often requires inspecting a large number of samples, which can be time-consuming. Instead, users should be able to quickly gain insights into a subgroup's content and relevance to detected errors. This could include textual summaries, representative samples, and relevant statistics.

\item[G3.] \textbf{Enable queries for specific concepts or aggressors.} CVML models may fail on samples containing certain ``semantic concepts'' or aggressors (e.g., \textit{people wearing hats}, \textit{mixed food}) that are often not pre-annotated. Users should be able to query and analyze samples associated with these concepts, thereby creating user-defined subgroups to validate whether they are linked to model errors.

\item[G4.] \textbf{Ensure easy access to subgroups throughout exploration.} The exploration of data subgroups can be highly iterative, often requiring comparisons between different data slices~\cite{cavallo2018clustrophile}. To facilitate this process, it is essential that the system tracks and provides easy access to explored subgroups for ongoing analysis.

\item[G5.] \textbf{Provide guidance for exploring relevant subgroups and samples.} In large datasets, users may benefit from guidance in deciding which subgroups or samples to explore. This could be achieved by pre-computing features of interest for each subgroup or by recommending subgroups that are semantically similar or perform similarly to the currently inspected one, helping users prioritize their exploration.

\item[G6.] \textbf{Offer a statistical basis to validate the impact of identified semantics on model performance.} When analyzing subgroup performance, MLEs should be able to compare the performance distribution of those samples against other subgroups and the overall dataset. This helps prevent users from prematurely applying assumptions about error causes to model development without first confirming that their observations generalize across the dataset and are not based on spurious correlations.

\end{itemize}
\begin{figure*}
    \centering
    \includegraphics[width=\textwidth]{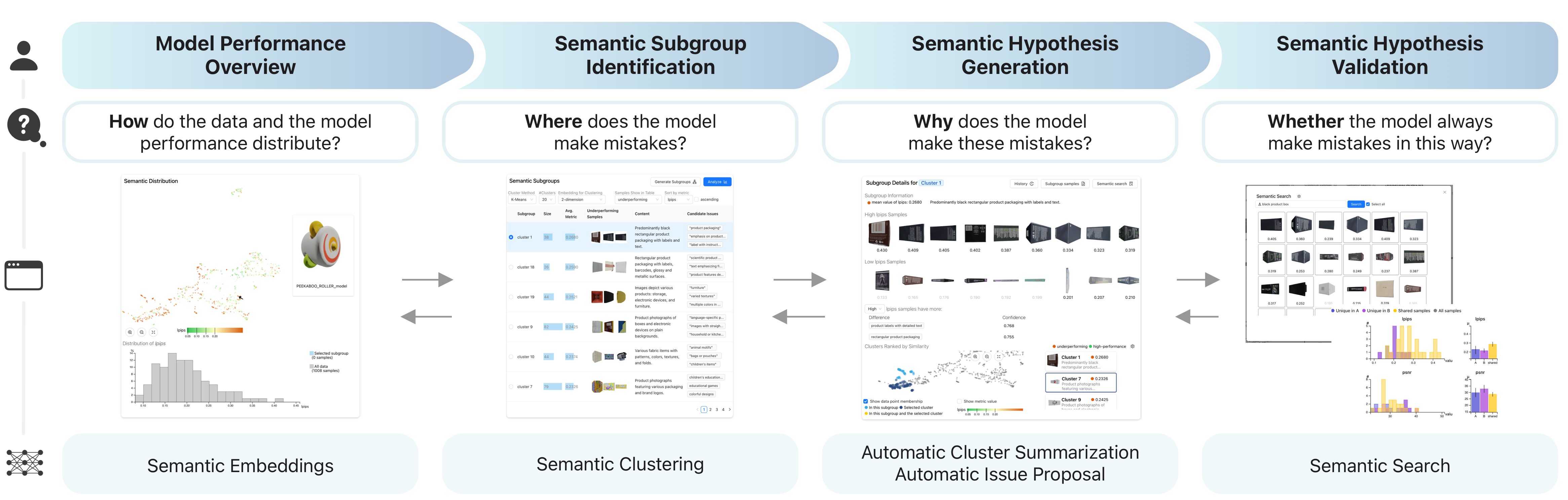}
    \caption{Semantic error analysis workflow with four main stages. Each stage aims to answer different questions, which is supported by different views in the user interface. To answer each question, we combine foundation models with visual analytics techniques.}
    \label{fig:workflow}
\end{figure*}

\section{VibE: Proposed Analytical Workflow and Methodology}
\label{sec:method}

Building on the formative study and design goals from the previous section, we present \systemname, an intelligent workflow for performing semantic error analysis at the subgroup level for CVML models. This section outlines key steps of the workflow and discusses the analytical techniques applied at each stage. The practical implementation is detailed in Sec.~\ref{sec:system}.

Our proposed workflow is data-centric, emphasizing the need to explore low performance within smaller semantic neighborhoods before verifying observations across subgroups and the entire dataset. It comprises four main stages: \textit{Model Performance Overview}, \textit{Semantic Subgroup Identification}, \textit{Semantic Hypothesis Generation}, and \textit{Semantic Hypothesis Validation}. As illustrated in Fig.~\ref{fig:workflow}, each stage supports users [\includegraphics[width=10pt, height=10pt]{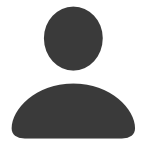}] in answering key questions [\includegraphics[width=10pt, height=10pt]{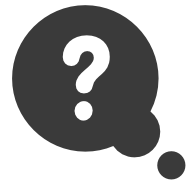}] through interaction with the user interface [\includegraphics[width=10pt, height=10pt]{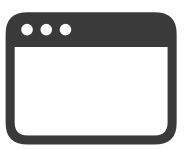}] of \systemname. The process is enhanced by large foundation models [\includegraphics[width=10pt, height=10pt]{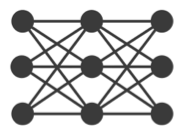}], and notably, the workflow is iterative, allowing users to independently revisit any stage as needed.

\begin{figure*}
	\centering
    \includegraphics[width=\linewidth]{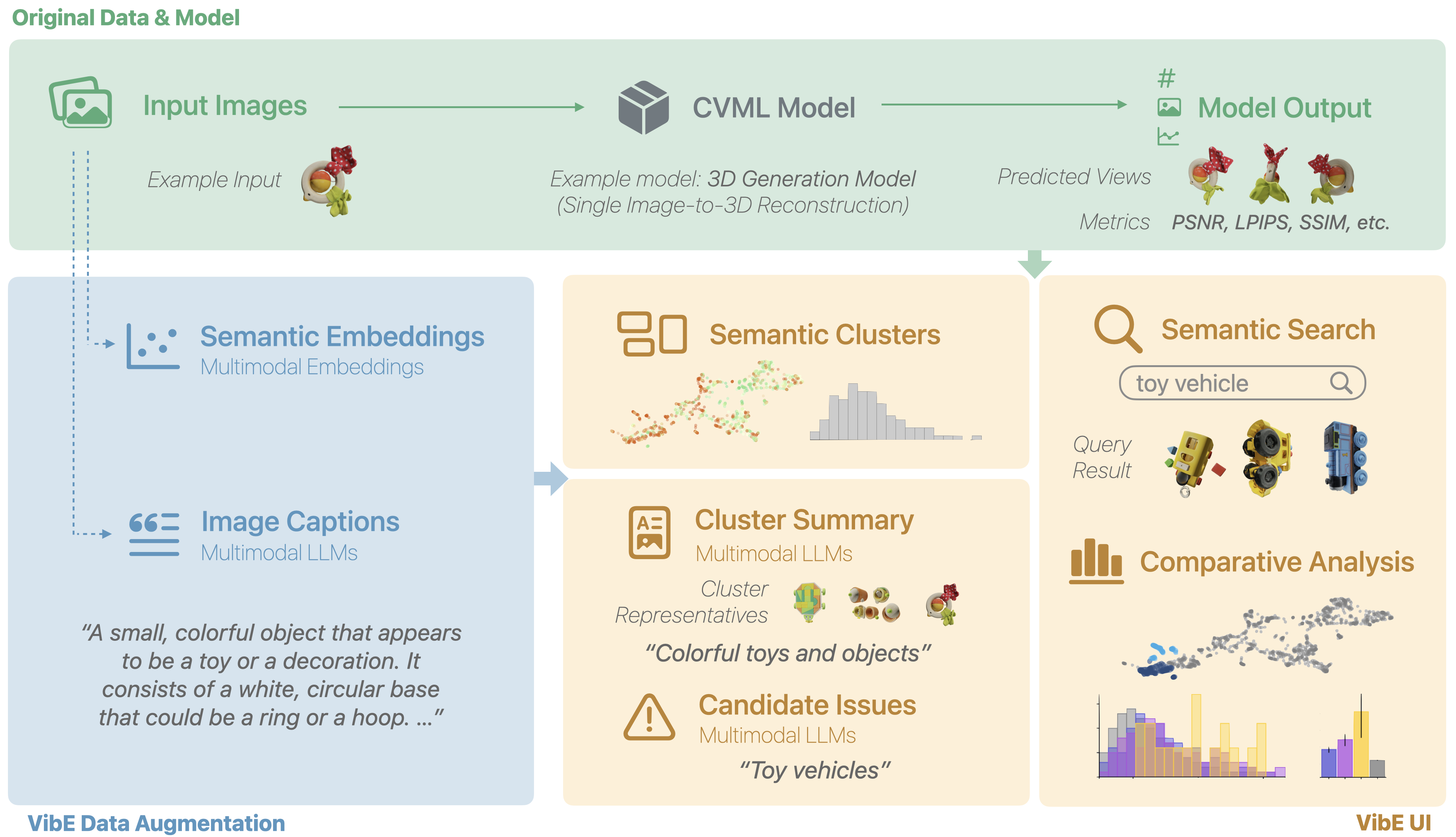}
  % \vspace{-20pt}	
   \caption{The data flow of \systemname{}. Based on the \dataflowgreen{original model input/output}, we generate \dataflowblue{auxiliary data} about the semantic meanings of model input images. These \dataflowblue{auxiliary data} facilitates the construction of  \dataflowblue{semantic clusters} and samples retrieval of \dataflowblue{semantic concept}. Both the \dataflowgreen{original model/data} and \dataflowblue{auxiliary data} are presented in the \dataflowyellow{Vibe UI}. \revise{In practice, the foundation models used in this workflow can be easily replaced with any model that aligns with users' preferences, data compliance needs, or other requirements.}}
  	\label{fig:data2ui}
\end{figure*}

\revise{We illustrate how data are processed by large foundation models and integrated into the user interface of \systemname{} in Fig.~\ref{fig:data2ui}. While specific models were selected during the implementation of \systemname{}, alternative models and algorithms can be chosen to suit tasks in different domains. Acknowledging the potential inconsistencies and biases of large language models (LLMs), we observed how users interacted with the information generated by these models and gathered their feedback through interviews, which we discuss in detail in the Discussion section.}

\subsection{Model Performance Overview}
\label{sec:stage_1}

The first stage of our workflow serves as a default starting point that can be integrated into most model evaluation tools, providing an initial overview of model performance through absolute metrics and data distribution patterns. This stage allows for basic filtering of data samples based on available metadata, enables ranking by metric values, and facilitates easy inspection of individual samples. Additionally, a visualization of the data's ``shape'' using dimensionality reduction (DR) techniques can reveal structures by clustering similar samples together.

We leverage the vector representations (embeddings) learned from widely available foundation models. In this work, these embeddings are derived from CLIP~\cite{radford2021clip}, a vision-language model trained contrastively on a large dataset of image-text pairs. CLIP’s text and image encoders map textual inputs $X_{t}$ and images $X_i$ to a latent representation $\phi \in R^{512}$ in a shared embedding space $E$. After obtaining $\phi$ for each image, these vector representations can be projected into a lower-dimensional space using DR techniques like UMAP~\cite{mcinnes2018umap} and displayed in interactive 2D scatterplots. While some high-dimensional information may be lost during DR, the 2D projection still preserves similarity patterns between samples and their neighbors. This projection can be conveniently combined with performance metrics, offering insights into correlations between semantics and model performance. For instance, we may observe that semantically similar samples (i.e., neighboring points) exhibit comparable performance trends. We select CLIP as a representative of any large foundational model capable of generating semantically meaningful embeddings. As discussed later (Sec.~\ref{sec:limit}), it may be beneficial to explore alternatives that are more specifically tailored to the CVML tasks being debugged.

\subsection{Semantic Subgroup Identification}
\label{sec:stage_2}
\label{sec:cluster_generation}
\label{sec:semantic_search}

Following initial analyses, we provide users with an initial set of candidate subgroups to explore (\textbf{G1}). To achieve this, we use the CLIP vector representations, combined with clustering techniques such as K-Means~\cite{sinaga2020unsupervised} and DBSCAN~\cite{schubert2017dbscan}. Operating within the semantically meaningful CLIP embedding space $E$~\cite{derby2018using}, all samples in the same cluster are considered to share semantic similarities. This clustering generates conceptually coherent subgroups, serving as a starting point for further exploration and analysis of model performance within distinct semantic neighborhoods.

The use of dimensionality reduction (DR) together with clustering techniques introduces several key trade-offs. Clustering in high-dimensional space (i.e., using the full 512-dimensional CLIP vectors) preserves all semantic information but suffers from the ``curse of dimensionality'', which increases computational costs and challenges interpretability. Conversely, clustering after applying DR techniques is faster and more visually interpretable~\cite{van2008visualizing, mcinnes2018umap}, though it offers only approximate similarity. DR methods reduce noise and focus on the most relevant features but may introduce unexpected biases or lose important details. Thus, the choice of technique significantly affects the resulting clusters, potentially leading to the loss of important semantic relationships during reduction. While these considerations are highly relevant, we limit our scope here and instead provide users with flexibility to adjust DR and clustering parameters, allowing them to balance between accuracy and efficiency based on their specific analysis needs.

Beyond an initial set of possible subgroups, we also support user-defined custom subgroups using the same CLIP embeddings (\textbf{G3}). For instance, a user can identify a subgroup by “querying” specific concepts through text inputs — a task that we refer to as ``semantic search''. Given a dataset $D$ containing $N$ image samples and a query $Q$, we return $n$ samples that satisfy the semantic concept described by our query $Q$.
We embed all database images with the CLIP Image Encoder $E$ into 512-dimensional latent representations $\phi$ and store them in a vector database before querying. During query time, a textual $Q_t$ query is encoded into $\phi_q$ using the CLIP text encoder. We then perform a similarity search based on the cosine similarity between the query vector $\phi_q$ and any database vectors $\phi_i$ as described below:

\begin{equation}
  \label{eq:distance}
  d\left(\phi_q, \phi_i\right) = \frac{\phi_q^T \phi_i}{\Vert \phi_q \Vert \Vert \phi_i \Vert}
\end{equation}

The $n$ closest samples according to the distance metric above are returned and displayed in the UI. In this way, we assist users in retrieving a subgroup of samples with the semantic features they have in mind. 

\subsection{Semantic Hypothesis Generation}
\label{sec:stage_3}

When visually inspecting samples within an identified subgroup, users can often infer shared semantic characteristics and relate them to individual metrics or the subgroup's overall performance. However, as the number of samples increases or semantic features become more nuanced, exhaustive inspection becomes impractical. To address this, we offer techniques to identify key features associated with each subgroup.

\subsubsection{Display of Subgroup Sample Representatives}
Selecting one or more samples to represent a subgroup is useful for various visual analytics tasks~\cite{molnar2022}. We can determine the centroid of each cluster by calculating the mean (or median) of the high-dimensional vectors corresponding to its samples. Then, by applying cosine similarity (Eq~\ref{eq:distance}), we identify the top $n$ samples closest to the centroid. These samples serve as effective representatives for the cluster (\textbf{G2}), particularly in interfaces with limited space or in visualizations like scatterplots, allowing quick reference to subgroups while minimizing visual clutter.

\begin{figure}
	\centering
  	\includegraphics[width=\linewidth]{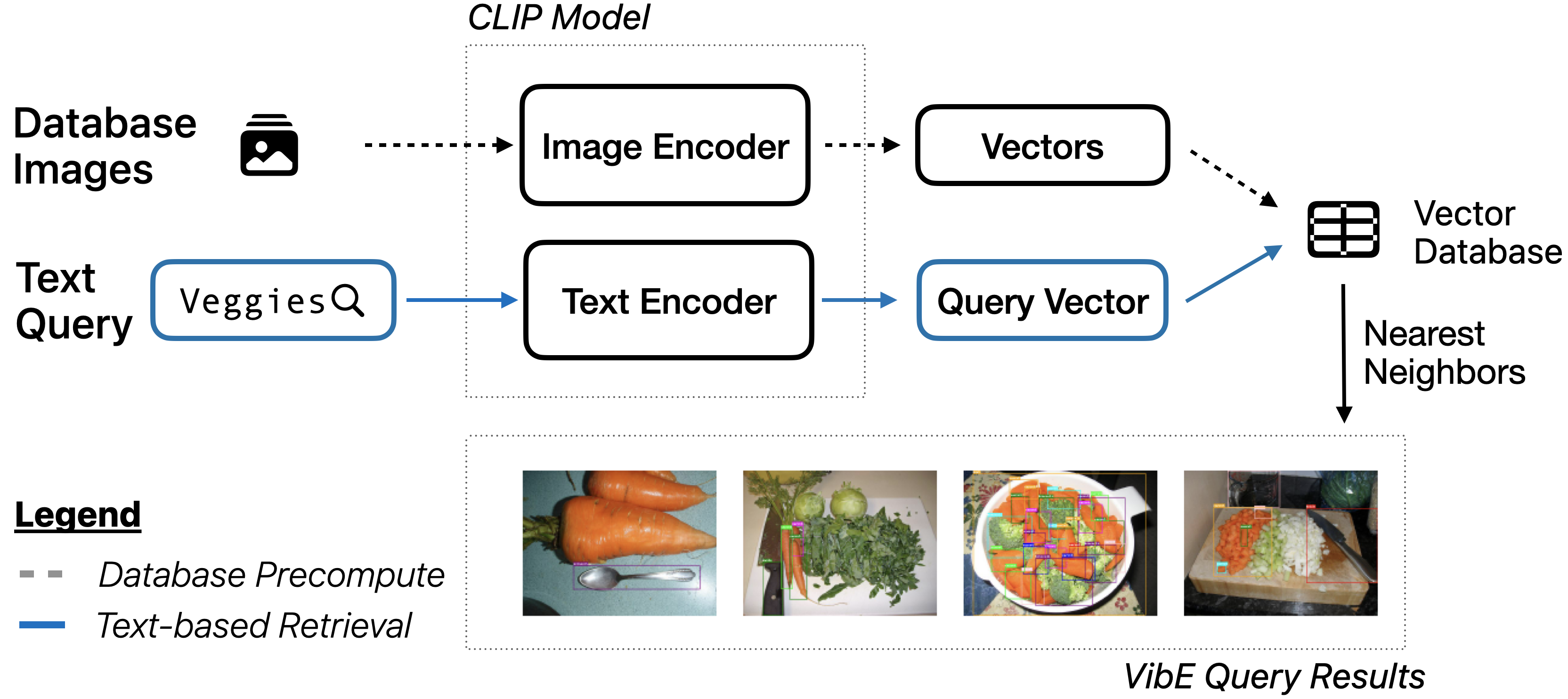}
   \caption{\revise{Image retrieval using natural language.} The retrieval pipeline uses a pre-trained CLIP model to retrieve image samples based on a text query.}
  	\label{fig:retrieval_explainer}
\end{figure}

Depending on the dimensionality reduction (DR) and clustering methods applied, representatives can be refined by filtering out outliers before computing the centroid. This ensures that the centroid accurately reflects the core semantics of the subgroup, rather than being skewed by samples that do not align with the group's general characteristics. We also explored synthesizing a representative image from the high-dimensional vector representations of the centroid through a diffusion process~\cite{podell2023sdxl}; however, this approach was not always reliable in accurately representing the shared characteristics of a subgroup.

While centroid-based representative samples capture overall subgroup semantics, performance analysis can benefit from displaying the lowest and highest performing samples within the cluster. By inspecting these extremes, users can identify recurring issues in low-performance samples and, when combined with metric distributions, observe whether hypothetical aggressors affect all samples or just a subset.

\subsubsection{Summarization of subgroup contents.}
\label{sec:keywords_generation}

As described in our formative interviews, understanding subgroups can be challenging due to the complex nature of data samples. We propose textual summarization as a powerful tool for describing subgroup semantics (\textbf{G2}). In our work, we use GPT-4 (\texttt{gpt-4o})~\cite{GPT4}, which supports various language and image variants. When sufficient ground-truth labels or annotations are available, we embed this metadata into a readable string format for the model's context. We then prompt the large language model (LLM) to describe commonalities within the provided data samples. An example template might be:

\begin{quote} ``I have the following data: {data}. Please summarize all these data using less than <num\_word> words. '' \end{quote}

Although GPT4o is able to take visual/image input, it is faster for it to generate response based on text only queries. In addition, many LLMs are hosted on the cloud, which requires sending the data to a third-party server. This might have data safety issues if the data contain personal identifiable information (PII) or other sensitive information. As such, we propose to summarize images based on the textual description or captions of them. 

However, most datasets lack sufficient metadata or annotations. To address this, we utilize another large language model to pre-generate textual captions for all images in the dataset. \revise{At the time of implementing \systemname{}, we selected 
LLaVA-NeXT~\cite{liu2024llavanext} for caption generation for two key reasons. First, it produces detailed image descriptions that support subsequent summarization tasks. Second, the model can be run locally, eliminating the need to send data to third-party services and thereby preserving any potential PII within the dataset.}
We then store these captions in a vector database and use GPT-4 to describe their shared characteristics, following the previous approach.

During our experiments with cluster summarization, a limiting factor was the LLM context size and the need for dataset-specific prompt tuning. If metadata or captions are too lengthy, or if we summarize too many samples, the prompt can exceed the model's context window. To mitigate this, we select a subset of samples based on proximity to the cluster centroid until we fill the available context length. We found that slight variations in prompt templates tailored to the specific use case yield better results. For instance, in the face reconstruction case presented in Section~\ref{sec:case_face}, we added in the prompt some facial characteristics examples that a user would like to see: ``The common characteristics can be related to gender, ethnicity groups, settings, facial features, outfits, events, image styles.'' Please find the detailed prompt content in our supplementary materials.

Subgroup summaries can also vary in length depending on the analytical task. For our system, we propose medium-length summaries (up to 15 words) for comparative tables, and longer descriptions for detailed single subgroup inspections. 

\subsubsection{\revise{Display and Ranking} of candidate semantic issues.}
\label{sec:candidate_issues}
Revisiting our previous example of highlighting low-performing samples within a subgroup, we expand the summarization concept to identify potential factors influencing model performance. Inspired by VisDiff~\cite{VisDiff}, we prompt GPT-4 to delineate the visual differences between high-performing and low-performing samples based on their captions, generating a list of possible ``aggressors'' expressed as 1-5 word concepts. We use the following template:

\begin{quote}
``The following are the result of captioning two groups of images: <caption placeholder>.

I am a machine learning researcher trying to figure out the major differences between these two groups so I can better understand my data.

Come up with 10 distinct concepts that are more likely to be true for Group A compared to Group B. Please write a list of captions. ''
\end{quote}

Here, \textit{<caption placeholder>} includes captions from 10 images in Group A and 10 in Group B, formatted as follows: ``Group A: <caption A1>; Group A: <caption A2>; ... Group B: <caption B1>; ...''. Group A contains the samples with the worst performance in a cluster or concept-based subgroup, while Group B consists of the best-performing ones. The prompt also specifies the correct and incorrect output formats. Please refer to our supplementary material for the complete prompt content.

These candidate issues highlight distinguishing characteristics associated with performance within the cluster and can be validated by visually inspecting the corresponding low-performing samples (\textbf{G2}). However, it is important to note that these results can only be computed if there is sufficient performance variability within the semantic cluster. If all samples exhibit uniformly high performance, generating a meaningful set of candidate issues becomes challenging. In contrast, if all samples have low performance, the cluster summary itself is a candidate issue of the this cluster.

To further assist users to identify the most relevant issues, we ranked all the proposed candidate issues by a \textit{confidence} score adapted from \textit{AUROC}(Area Under the Receiver Operating Characteristic Curve) score for a classification model. More specifically, given a GPT-generated candidate issue text as $Q_t$ and its 512-dimensional CLIP embedding as $\phi_q$, an image in either Group A or Group B as $M_i$ and its CLIP embedding as $\phi_i$. We can then calculate a cosine similarity $d(\phi_q, \phi_i)$ between the candidate issue and each of the images in Group A or B using equation~\ref{eq:distance}. Then we use $d(\phi_q, \phi_i)$ as a probability of whether an image belongs to Group A (underperforming samples) or not. In this way, we have a simple classifier to differentiate Group A and Group B after setting a probability threshold (i.e., if $ d(\phi_q, \phi_i)>0.5$, then it is from Group A). Here we use the performance of this simple classifier as a proxy of whether a candidate issue can well differentiate Group A (underperforming samples) from Group B (well-performed samples). Because \textit{AUROC} can evaluate a classifier's performance across all the possible thresholds, we use it as the \textit{confidence} score of a candidate issue.

\subsection{Semantic Hypothesis Validation}
\label{sec:stage_4}
The visual inspection of a subgroup's relevant samples, along with system-generated suggestions, can prompt machine learning engineers (MLEs) to identify initial issues affecting model performance. Before modifying the algorithm or data, it is crucial to iteratively refine these observations and validate them through basic statistical analysis. After examining a subgroup of interest and formulating potential hypotheses, MLEs should verify their findings in similar subgroups and across the entire dataset.

\subsubsection{Guidance in exploring similar subgroups.} 
Using CLIP embeddings, we can leverage cluster centroid positions and their distances from the currently considered subgroup (as described in Equation~\ref{eq:distance}) to identify semantically similar clusters (\textbf{G5}). If neighboring clusters exhibit comparably low performance, inspecting them may reveal shared issues. Conversely, high-performing neighboring clusters may indicate that the hypothesized problems do not generalize or that the semantic features examined are too specific and not accurately represented by the current embedding, dimensionality reduction (DR), or clustering methods. Further verification can be achieved by adjusting these parameters or conducting dedicated searches. Additionally, neighboring subgroups that display mixed performance may uncover hidden aggressors for further validation.

\subsubsection{Semantic (Concept) Search}
After initial iterations, MLEs can refine their observations to one or more plausible semantic issues and use the semantic search functionality from Sec.\ref{sec:semantic_search} to generate a new subgroup based on those concepts (\textbf{G3}). If samples consistently exhibit low performance, it may indicate the concept represents an aggressor. Conversely, if performance is mixed, the new subgroup can serve as a starting point for further analysis. Examining high and low-performing samples and associated candidate issues may reveal nuanced semantics and lead back to the hypothesis generation step in Sec.\ref{sec:stage_3}. Even clusters with solely low-performing samples can benefit from additional semantic search queries to filter out spurious correlations and ensure model failures are not misattributed to other coinciding semantic characteristics.

\subsubsection{Subgroup Analysis}
At this stage, MLEs often iterate through multiple subgroups, comparing them to see if observations apply to others or generalize across the dataset. To strengthen these comparisons, visual inspection should be paired with basic statistical assessments: Do the subgroups have enough samples for meaningful observations? How do their performance metric distributions compare, and are they significantly different from the overall dataset? When comparing a system-generated cluster with a user-generated subgroup (e.g., based on a specific concept), how many samples overlap? These questions aim to evaluate the generalizability and significance of generated hypotheses quantitatively (\textbf{G6}).

As noted by Dragicevic~\cite{dragicevic2016fair}, it would be beneficial to report informative charts and interval estimates with explanations when comparing two groups. We include histograms to illustrate the shape of the metric distribution for one or two subgroups. Additionally, we calculate each subgroup's mean value as a summary statistic, while the $95\%$ bootstrap confidence interval (CI) indicates the precision and reliability of the mean estimate. If the CIs of two subgroups do not overlap, this suggests a significant difference between their mean values. Conversely, significant overlap may imply that apparent differences in means could result from random variation. A narrow CI indicates high precision in the mean estimate, while a wide CI suggests uncertainty or variability in the data. Together, these methods provide complementary insights for comparing two subgroups.

\section{\systemname: System Implementation}
\label{sec:system}
In this section, we present the implementation of the analytical workflow introduced in Sec.\ref{sec:method}. This implementation not only demonstrates how the proposed methodologies can be integrated into a fully functional system, but also enables us to evaluate their effectiveness in Sec.\ref{sec:study}.
Our \systemname{} implementation was developed through an iterative design process, with our developers collaborating biweekly with a UI/UX designer and a machine learning engineer over twelve months to refine the system design and validate its utility. The system is built using a common software stack, featuring a React-based web interface and a Python backend that handles data loading and computation, supported by a local PostgreSQL database with the pgvector extension.
Leveraging focus-plus-context visualization techniques~\cite{focus_plus_context}, the system allows users to explore subgroups in detail while maintaining an overview of the entire dataset. This enables iterative hypothesis generation and validation related to error causes. Below, we describe the two main pages of the system and their respective views.

\begin{figure*}
    \centering
     \includegraphics[width=\linewidth]{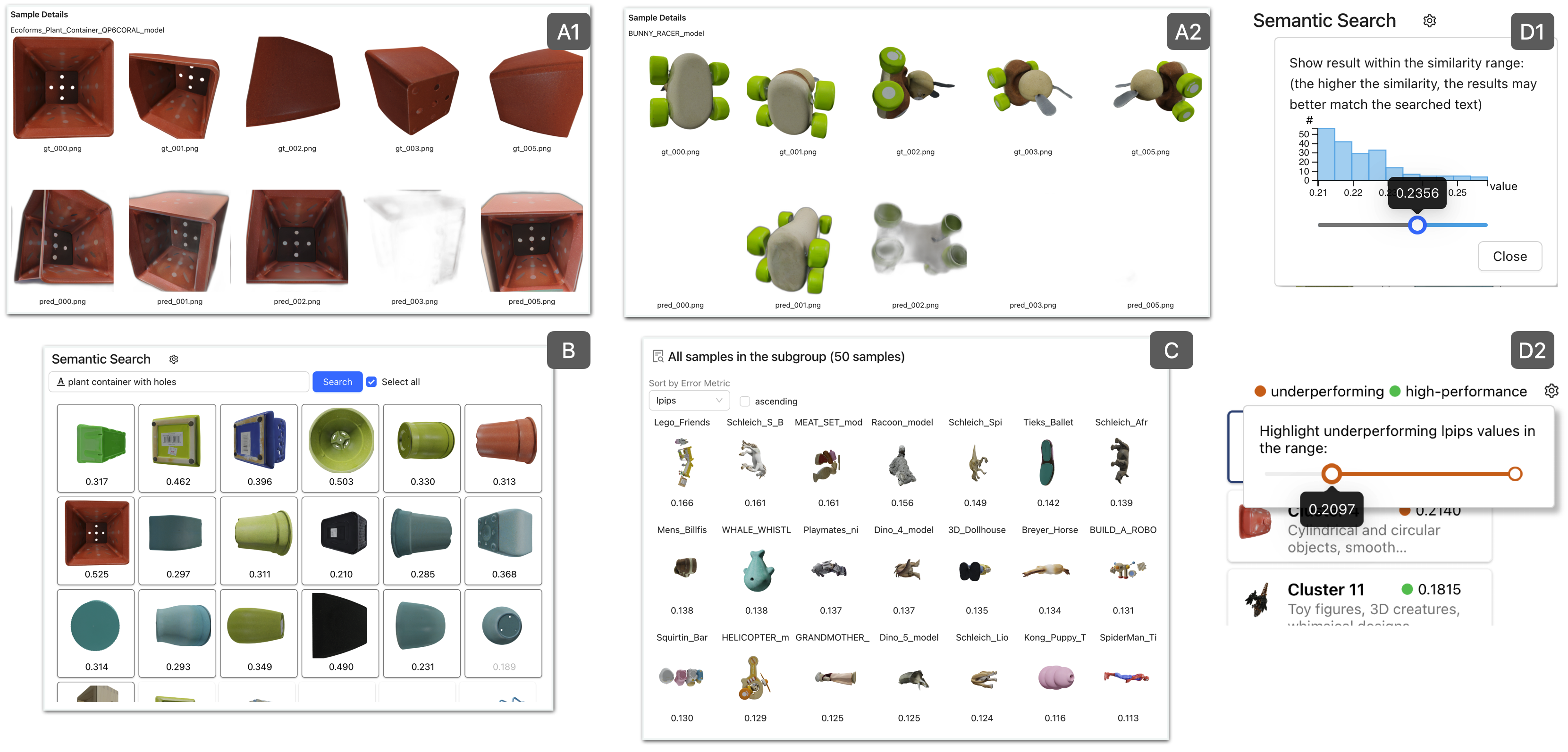}
    \caption{{\systemname} provides modal windows to support users to (A) check detailed model input and output, (B) retrieve samples by semantic concept search, and (C) inspect all the samples in one subgroup ranked by specific metric value. \revise{(A2) illustrates an example where the model failed to generate views from certain angles.} {\systemname} also provides some setting functions to assist users to (D1) filter the semantic search results and (D2) customize the error color encoding.}
    \label{fig:supporting_3d}
\end{figure*}

\subsection{Dataset Overview}
The initial overview page (Fig.~\ref{fig:ui_3d}, A) predominantly corresponds to the first two stages of our workflow (Sec.\ref{sec:stage_1} and Sec.\ref{sec:stage_2}), and serves as a starting point for users, similar to how other tools might begin their subgroup analysis. It features commonly used visualizations, such as a dimensionality-reduction-based scatterplot and a performance distribution chart, along with initial tools for generating candidate subgroups for exploration.

\subsubsection{Semantic Distribution Panel}
This panel includes a histogram that displays the distribution of a selected performance metric across all dataset samples (Fig.~\ref{fig:ui_3d}, A1), paired with an interactive 2D scatterplot designed to reveal structures with shared semantic characteristics (Fig.~\ref{fig:ui_3d}, A2). By default, the scatterplot utilizes UMAP~\cite{mcinnes2018umap} to reduce high-dimensional CLIP representations into two dimensions. However, users can switch to other techniques like t-SNE, MDS, and PCA, in line with previous works (~\cite{cavallo2018clustrophile, nonato2018multidimensional,wenskovitch2017towards}). Scatterplot points are color-coded based on their performance metric value, allowing similarly colored neighborhoods to highlight potential correlations between semantic similarity and model performance.

\subsubsection{Semantic Subgroups Panel}
The semantic subgroups panel (Fig.~\ref{fig:ui_3d}, A3) provides an initial set of candidate subgroups for analysis. Users can configure various parameters to define these subgroups, such as the clustering method, the number of clusters (if applicable), and the type of computation (high vs low-dimensional embeddings). Clusters can be ranked by an aggregate performance metric, offering a prioritized view of those potentially worth deeper inspection.
The candidate subgroups are displayed in a table, with each row showing the absolute and relative size, as well as the average value of the selected performance metric for each cluster. To enhance the user experience, we integrate techniques like cluster representatives and summaries, as discussed in Sec.~\ref{sec:stage_3}, and cache their results. This provides upfront context as to why a specific cluster might warrant attention.
In particular, each cluster displays semantically representative samples and a textual summary describing its contents. Additionally, low-performing samples and potential causes (i.e., ``aggressors'') are highlighted to give users initial insights into possible performance issues. Upon selecting a cluster, users can view its metric distribution (histogram in blue) compared to the overall dataset (histogram in gray). Cluster samples are also highlighted in the scatterplot, supporting users in determining whether the samples are outliers or belong to a larger semantic or similarly performing group. Users then have the option further inspect a specific cluster by transitioning to the Subgroup Analysis page (Fig.~\ref{fig:ui_3d}, B).
% Clicking \lookupbtn{} opens a modal window of sample overview (Fig.~\ref{fig:modal}b), where users can take a look at all the samples ordered by metric of interest in an ascending or descending order.

\subsection{Subgroup Analysis}
This second page focuses on a more detailed inspection of a single cluster and of its neighbors (Sec.~\ref{sec:stage_3}), and provides some statistical analysis tools to verify error cause hypotheses (Sec.~\ref{sec:stage_4}). Semantic search further provides a way to iteratively move from detail to overview, enabling the generation of conceptual subgroups across the dataset.

\subsubsection{Subgroup Details Panel}
When inspecting a single cluster, we display the textual summary and the top high and low performing samples as described in (Sec.~\ref{sec:stage_3}). While a button allows opening a modal to inspect all samples within a subgroup \revise{(Fig.~\ref{fig:supporting_3d}, C)}, these samples at the tails of the distribution are often enough to get a sense of its contents (Fig.~\ref{fig:ui_3d}, B1). Low-performing samples can also be visually matched with the list of candidate error issues computed with GPT-4 (Sec.~\ref{sec:candidate_issues}) and displayed below (Fig.~\ref{fig:ui_3d}, B2). The setting further allows to highlight only samples within a given range of performance metric values, which are displayed under each sample. While we display both ground truth and prediction, if applicable, for each sample, a dedicated debugging modal is provided to display any additional file or custom visualization associated with a sample. For instance, in our 3D asset generation case study (Sec.~\ref{sec:case_3dgen}) we display employ this modal to display multiple ground truth / prediction couples associated with different camera angles.

% \subsubsection{Neighboring Clusters Panel}
In the same panel, we offer a display of clusters neighboring the current subgroup (Fig.~\ref{fig:ui_3d}, B3). Each neighboring cluster is represented as a card that includes a representative sample, a short textual description of its contents, and an indication of whether its aggregate performance is high or low, based on the selected metric and the user-defined value range. This is paired with the scatterplot visualization, allowing users to observe whether semantically similar clusters exhibit comparable performance. Such similarities may suggest the generalizability of an identified issue, while clusters that mix high- and low-performing samples might warrant further investigation to detect possibly hidden aggressors. Additionally, color coding for high and low metric values can be inverted in cases where lower values correspond to higher errors.

\subsubsection{Semantic Search Modal}
After identifying potential semantic features linked to low performance within a cluster and its neighboring clusters, users can open the CLIP-enabled semantic search modal to query these concepts across the entire dataset. This allows them to form a new, custom subgroup to assess whether those characteristics are more broadly associated with low performance (Sec.~\ref{sec:stage_4}). The previously mentioned metric value threshold can be applied to refine search results before finalizing the creation of the new subgroup.
The Semantic Search modal can also be opened by directly clicking on the aggressors listed in the Subgroup Details panel.

\subsubsection{History Modal}
Given the iterative nature of data exploration tasks~\cite{cavallo2018clustrophile}, we provide a modal listing all previously inspected clusters and subgroups generated via semantic search (\textbf{G4}), using the same card-based representation for each subgroup. Users can quickly reload any subgroup into the Subgroup Details Panel by clicking on its card. Additionally, subgroups can be selected from this list for comparison in the Subgroup Analysis panel, as described below.

\subsubsection{Subgroup Analysis Panel}
Users can select one or more subgroups from the History Modal for analysis in a dedicated comparative statistics panel (Fig.\ref{fig:ui_3d}, B4-B5). This panel juxtaposes the selected subgroups with bar charts that illustrate their absolute and relative sizes, comparing them to the entire dataset and highlighting the number of shared samples (Fig.\ref{fig:ui_3d}, B4). Relevant settings are accessible through checkboxes in the UI.
In the lower half of the panel, subgroup comparisons shift to the perspective of metric distributions, considering all available metrics simultaneously (Fig.~\ref{fig:ui_3d}, B5). Subgroup distributions are presented as differently colored histograms, with shared samples excluded by default to emphasize the differences among the selected clusters. Additionally, the same comparison is provided in aggregate form through a bar chart that displays the average values and $95\%$ confidence intervals for each metric across subgroups. To assist users in interpreting the results, we include an optional textual explanation indicating whether the performance differences between two subgroups or between a subgroup and the entire dataset are statistically significant, based on the overlap of these confidence intervals (Sec.~\ref{sec:stage_4}).

\section{Case Studies}
\label{sec:study}
In this section, we qualitatively evaluate our {\systemname} workflow and methodology by exploring three case studies centered around three diverse tasks,
% 3D asset generation, face reconstruction, and object detection. 
\revise{ranging from a traditional CVML task, object detection, to a generative model, 3D asset generation, as well as an important face reconstruction task that involves model fairness.}
Our study involved a total of 6 machine learning engineers (E1-6, 2 female, 4 male, all aged from 25 to 35) from a major technology company, with 2 engineers using our {\systemname} system implementation to address each of the three CVML tasks. The participants chosen for each task declared to have 2-5 years of work experience developing and debugging CVML models for that particular task. For each task, we chose to use public datasets and state-of-the-art model architectures that closely resemble the real-world challenges faced by the participants.

Instead of focusing on evaluating our implementation, our main interest revolves around understanding how model developers would employ our proposed methodologies to analyze their model's behavior, and how foundation models can assist them during the model debugging workflow. We particularly aim to investigate whether the auto-generated summaries and \revise{semantic} search based on large foundation models can help model developers discover new insights, and come up with ways to improve their models. Ultimately, it is our goal to explore how foundation-model-based semantic error analysis workflows could shape future CVML error analysis practices.

\textbf{Procedure.} We began each interview with an introduction, during which we clarified the goal of {\systemname} and provided a tutorial of {\systemname} ($10$ minutes). Then, we asked participants to analyze the type of model they work closely with to figure out where and why the model makes mistakes. Specifically, E1 and E2 analyzed a 3D asset generation model; E3 and E4 explored the face reconstruction model; E5 and E6, the object detection model. During the process, we instructed the experts to follow a ``think-aloud'' protocol~\cite{fonteyn1993description}, where they reasoned out loud and explicitly mentioned what questions they were trying to answer during the exploration and what insights they gathered ($15-20$ minutes). In the end, we conducted a semi-structured interview that incorporated several questions about the overall usefulness of {\systemname}, the functions they liked most, those they did not use frequently, and how they utilized information from foundation models, such as subgroup summarization outputs and concept search results ($15-20$ minutes). \revise{We also asked participants how they compared {\systemname} with their current workflows and tools that did not leverage foundation models.} Each session was conducted in person. With participants’ consent, each session was recorded and transcribed for analysis. After the meeting, E3-E6 volunteered to offer their own data and model output to the authors and provided follow-up feedback after using \systemname{} with their own data.

In the following subsections, we describe how \systemname{} was used for the three CVML tasks by the users. \revise{The users' feedback led to a discussion, detailed in the next section, about the role of \systemname{}, how participants integrated it into their workflows, and our lessons learned about effectively combining foundation models and visual analytics to design intelligent user interfaces for complex data analysis tasks.}

\subsection{3D Asset Generation Model: Error Discovery and Interpretation}
\label{sec:case_3dgen}
Although we did not include model developers for 3D asset generation in our formative study, we recognized that {\systemname} could be beneficial for this use case based on insights gathered throughout the project. Therefore, we included 3D asset generation models as part of the evaluation of {\systemname}.

In this case study, we asked participants (E1, E2) to use the diffusion model Zero123++~\cite{shi2023zero123plus} on the Google Scanned Objects (GSO)~\cite{downs2022google}, a large-scale real-world dataset of 3D scanned household items, to generate novel views of a 3D object from different camera angles based on a single image of that object. These multiple views are then typically used to generate the geometry and texture of the 3D object through a separate process. For this study, we used the GSO validation set, which contains $1,030$ high-quality scanned objects. In this dataset, each object has a high-quality image from 56 different angles as ground truth. To prepare data for clustering and semantic search, we generate CLIP embeddings for all the 56 images of an object. And then store the mean vector of the 56 CLIP embeddings as the semantic vector for this object. We sent an image from a random angle to Zero123++ as an input, and then select model output from 6 different angles to measure the errors on this object.

To evaluate the performance of the model, developers relied on several metrics to measure the differences between the generated views and the ground truth views (scanned object views). For instance, the \textit{Learned Perceptual Image Patch Similarity (LPIPS)} metric evaluates the perceptual dissimilarity between image patches—higher \textit{LPIPS} values~\cite{zhang2018perceptual} indicate greater differences between the generated images and the ground truth. Additionally, we included metrics such as \textit{Peak Signal-to-Noise Ratio (PSNR)}\cite{huynh2008scope} and its variant \textit{PSNR-mask}\cite{ponomarenko2011modified}, along with \textit{Structural Similarity (SSIM)}~\cite{wang2004image} to measure model performance.

We detail below the workflow observed from participant E1 to showcase a full debugging session.

\textbf{Exploring error samples.}
Participant E1 began by examining the model's performance in the Semantic Distribution panel. Points in the projection view were colored by their performance, allowing underperforming samples to stand out visually. E1 hovered over these points to preview the model’s outputs, which gave them an initial understanding of the failing samples and the extent of errors. Through this exploration, E1 identified several regions with high error rates, seemingly consisting of colorful boxes and toys.

\textbf{Generating and exploring semantic clusters.}
E1 used the tool to generate 20 semantic clusters based on 2D vectors of CLIP embeddings. These clusters were ranked by their \textit{LPIPS} values, making it easier to spot clusters where the model underperformed. By selecting the generated clusters from Semantic Subgroups panel, E1 could highlight samples in the projection view, graying out unselected ones. E1 toggled between clusters and hovered over highlighted points to inspect specific samples, eventually identifying errors such as vases and containers, which were initially overlooked during random point checks.

\textbf{Analyzing samples in one cluster.}
Next, E1 selected \textit{Cluster 19}, which contained various containers, for further analysis. In the subgroup analysis view, E1 clicked the sample with the highest \textit{LPIPS} error to examine all generated views, discovering that the model struggled with the geometry of the container from multiple angles. Other high-error containers in this cluster exhibited similar issues with generating the correct shape.

\textbf{Searching for a specific concept.}
As suggested by the auto-generated visual differences (Fig.~\ref{fig:ui_3d}, B2), E1 hypothesized that the model struggled with ``objects with holes or cutouts.'' Searching for such samples, E1 found that few appeared in \textit{Cluster 19}. Upon closer inspection, E1 refined the search to ``plant container with hole'' (i.e., plant pots or planters with drainage holes), which revealed samples with the highest \textit{LPIPS} values in \textit{Cluster 19}.

\textbf{Analyzing samples across subgroups.} 
E1 wondered if this concept could explain the high-error samples in \textit{Cluster 19}. Comparing two subgroups—\textit{Cluster 19} and the concept \textit{``plant container with holes''}—E1 found that both had significantly higher average \textit{LPIPS} values than the entire dataset, suggesting error-related features in both subgroups. Highlighting shared samples (Fig.\ref{fig:ui_3d}, B4), E1 noted that 22 samples were retrieved from the user-defined concept, 7 of which overlapped with \textit{Cluster 19}. Further statistical analysis (Fig.\ref{fig:ui_3d}, B5) led E1 to conclude that the concept could be attributed to the higher errors in \textit{Cluster 19}. After removing these samples, the remaining ones no longer exhibited a significantly higher mean \textit{LPIPS} value compared to the whole dataset.

\textbf{Checking neighboring clusters with similar content.}
E1 didn’t stop at analyzing the retrieved concept in isolation. To investigate whether the model struggled with other types of containers, E1 compared the \textit{``plant container with holes''} concept with neighboring clusters in the projection view (Fig.~\ref{fig:ui_3d}, B3). This comparison helped E1 understand the distribution of samples containing the concept and identify other clusters with similar content. E1 then initiated the next round of analysis by inspecting \textit{Cluster 10}, which also contained similarly perforated containers and had a higher mean \textit{lpips} score.

\textbf{Variations in participant analysis strategies.}
E2 followed similar initial steps as E1 to gain an overview during the exploration. However, with more experience using both the GSO dataset and the Zero123++ model, E2 identified a different cause for errors. While examining the projection view, where many toy samples were clustered, E2 noticed a toy car-like sample viewed from the bottom. E2 immediately recognized that this sample should depict a toy car with a bunny on top. By checking the sample details and all available views (Fig.~\ref{fig:supporting_3d}, A2), it was confirmed that this sample indeed featured a bunny on the toy car. E2 explained to the authors that Zero123++ typically performs better with input images from specific angles; thus, the bottom view perspective of the bunny car did not yield favorable results.
Additionally, E2 utilized the statistical analysis in the subgroup comparison panel less frequently than E1. Instead, E2 focused on validating various concepts they suspected might hinder the model’s performance by examining the retrieved samples and their corresponding performance metrics.

\subsection{Face Reconstruction Model: Aggressor Analysis \& Fairness Test}
\label{sec:case_face}
Our second use case involves employing a Variational AutoEncoder (VAE) model for face reconstruction. This CVML task aims to learn a robust vector representation of each face image, which is often utilized for downstream tasks, ranging from exploratory data analysis to integration into other model architectures. In this case, we utilize an open-source VAE model~\cite{toledo2021face} trained on the CelebA dataset~\cite{liu2015faceattributes}, a large-scale collection of celebrity face attributes. We evaluate the model on the CelebA test set, which comprises $19,867$ human faces. The objective of this analysis is to assess how well the model preserves facial features in its synthetically generated output images across different demographics. This evaluation leverages several metrics, including Kullback–Leibler Divergence~\cite{csiszar1975divergence} (KLD) and Mean Squared Error (MSE). We also employ a loss function as a linear combination of these two metrics; thus, a higher loss indicates poorer model performance. 

\begin{figure*}
  \centering
  \includegraphics[width=\linewidth]{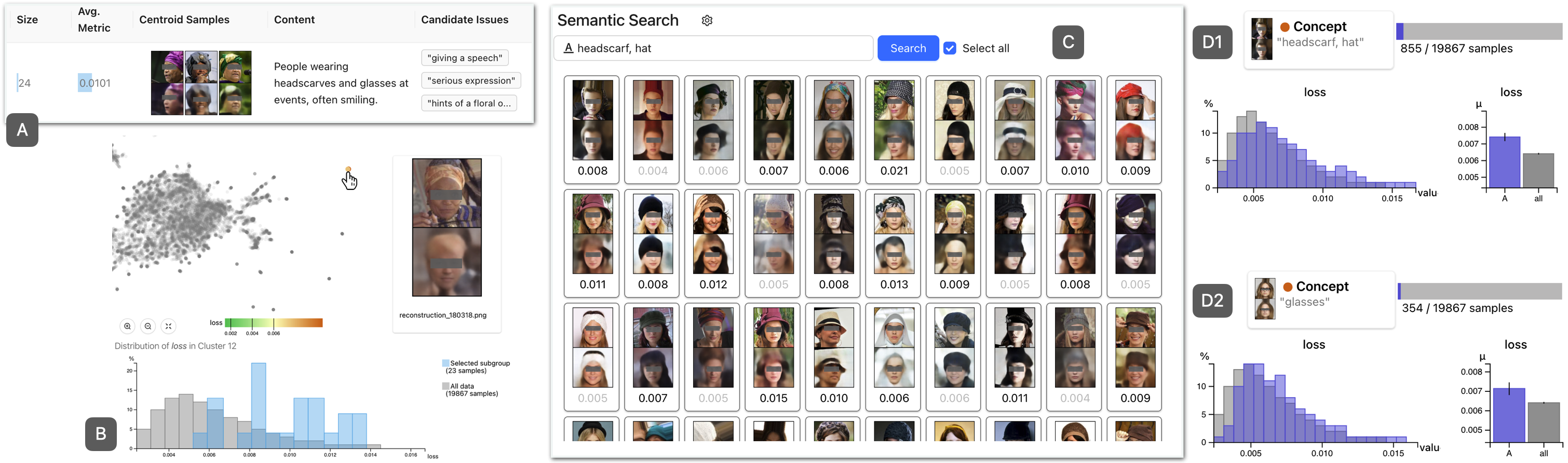}
  \caption{Using {\systemname} to analyze a face reconstruction model. For each image, the top half is the input image, the bottom half is the synthetically generated output by the model.
  }
  \label{fig:ui_face}
\end{figure*}

We outline below how E4 analyzed and reasoned about the problematic model output.

% check the aggressors
% check the distribution in different subpopulations 

\textbf{Data overview.} E4 began the exploration using the \textit{Semantic Distribution} panel, where they first examined the distribution of samples. E4 observed a general transition from male to female representations (top to bottom) and a gradient from lighter to darker skin tones (left to right). E4 continued to investigate the data points displayed in the projection view for additional insights.

\textbf{Cluster inspection and aggressor analysis.} E4 generated 50 clusters and focused on those with the highest \textit{loss} values. Notably, \textit{Cluster 12}, the cluster with the highest loss, appeared as an outlier, situated far from the majority of data points (see Fig.~\ref{fig:ui_3d}B). The cluster summary indicated that samples in this group predominantly featured individuals wearing headscarves and glasses. Upon further analysis, E4 discovered that this cluster primarily contained images of the same woman, wearing headscarves of various colors while delivering speeches at different events. This consistency among samples accounted for the poor performance and the tight clustering observed in the projection view.

\begin{figure}
    \centering
    \includegraphics[width=0.8\columnwidth]{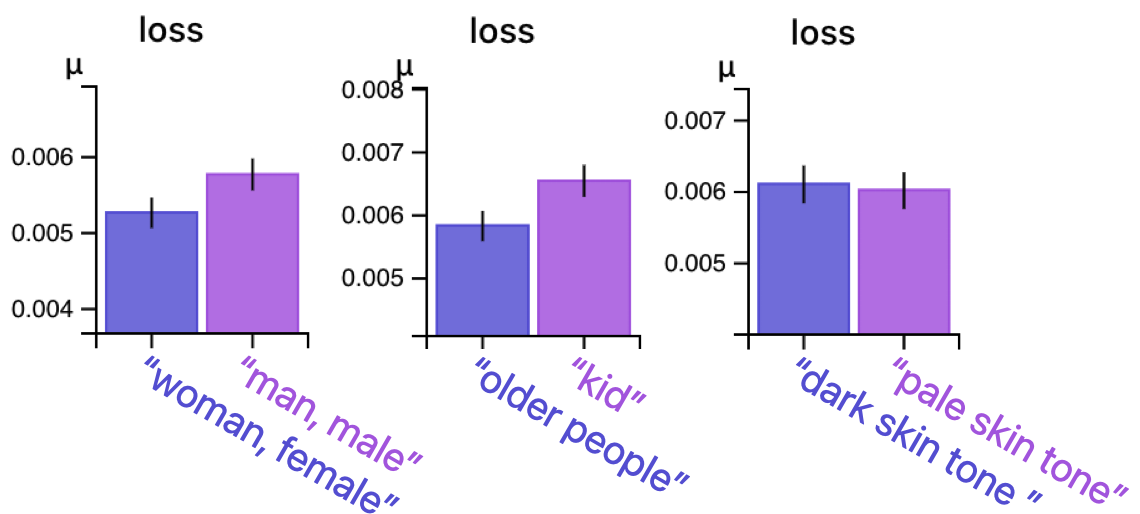}
    \caption{Users check whether a model has fair performance across different subpopulations by querying samples from semantic concepts related to gender, age, and skin tones. The first two charts show significant differences on loss while the last chart does not show significant difference.}
    \label{fig:face_fairness}
\end{figure}

E4 mentioned that in their daily work, the model is tested by focusing on human faces with specific ``aggressors''. These typically refer to inputs or features that cause significant or critical errors in the model’s predictions. An aggressor can also signify a systemic error in the model, where certain patterns or types of data consistently lead to failures. E4 aimed to analyze whether the model exhibited consistent mistakes with common aggressors.

E4 first searched for ``headscarf, hat'', as \textit{Cluster 12}, which featured a woman wearing a headscarf, had already demonstrated low performance. They then examined the metric distribution for the subgroup associated with the concept \textit{``headscarf, hat''} and compared it to the overall dataset. As shown in Fig.~\ref{fig:ui_3d}-D1, this subgroup exhibited a significantly higher average \textit{loss} value than the entire dataset.
Next, E4 followed similar steps to investigate the concept \textit{``glasses''}, which also displayed a notably higher average \textit{loss} compared to the full dataset (see Fig.~\ref{fig:ui_3d}, D2). After reviewing several high \textit{loss} samples related to these aggressor concepts, E4 confirmed that the VAE model struggled when human faces were partially obscured by decorations such as hats and glasses. This often resulted in blurred or incorrect representations of facial features (e.g., a red hat being reconstructed as red hair).

\textbf{Fairness test.} Based on prior experience, E4 aimed to ensure fair performance across different demographic subpopulations. To achieve this, E4 used the semantic search feature to retrieve and compare performance metrics for the following subgroups: ``man, male'' and ``woman, female'', ``kid'' and ``older people'', and ``dark skin tone'' and ``pale skin tone''. As shown in Figure~\ref{fig:face_fairness}, samples in the ``man, male'' subgroup exhibited significantly lower performance than those in the ``'woman, female'' one. E4 noted that the CelebA dataset contains many more female samples than male ones~\cite{quadrianto2019discovering}, which may explain the model's better performance with female faces.
In terms of age groups, E4 inspected several samples of kids and observed that the underperforming samples in the \textit{``kid''} category often featured fashionable decorations such as hats and sunglasses. In contrast, the samples of older individuals in the dataset were predominantly portrait-like images with well-defined hair shapes and facial features. This pattern echoed the earlier issues E4 identified with aggressors. Ultimately, E4 concluded that the model performed fairly well across different skin tones.

\textbf{Variations in participant analysis strategies.} E3 and E4 employed similar analysis strategies regarding aggressors and fairness tests. E3's analysis involved searching for various concepts such as ``blonde hair'', ``beard'', ``facial hair'', and ``afro''. Since they were working with a model focused on human faces, ensuring fair performance and mitigating the impact of aggressors were two critical aspects of their daily work.

\subsection{Object Detection Model: Inconsistent Annotations}
\label{sec:case_food}
Object detection involves identifying which objects, as defined by a predefined taxonomy, are present in an image and determining the regions they occupy. In this third case study, we apply the open-source YOLO model~\cite{yolo} to the validation set from the COCO~\cite{coco} dataset, which consists of 4,905 images featuring commonly encountered objects in our daily lives. The evaluation criteria for this study focus on three main aspects: \textit{detection} (whether the predicted bounding boxes align with the ground truth), \textit{classification} (whether the detected areas are classified correctly), and \textit{recognition} (whether both the predicted bounding box and the class match the ground truth).
We describe below how E5 analyzed and reasoned about the label and annotation issues within this use case.

\begin{figure*}
  \centering
  \includegraphics[width=\linewidth]{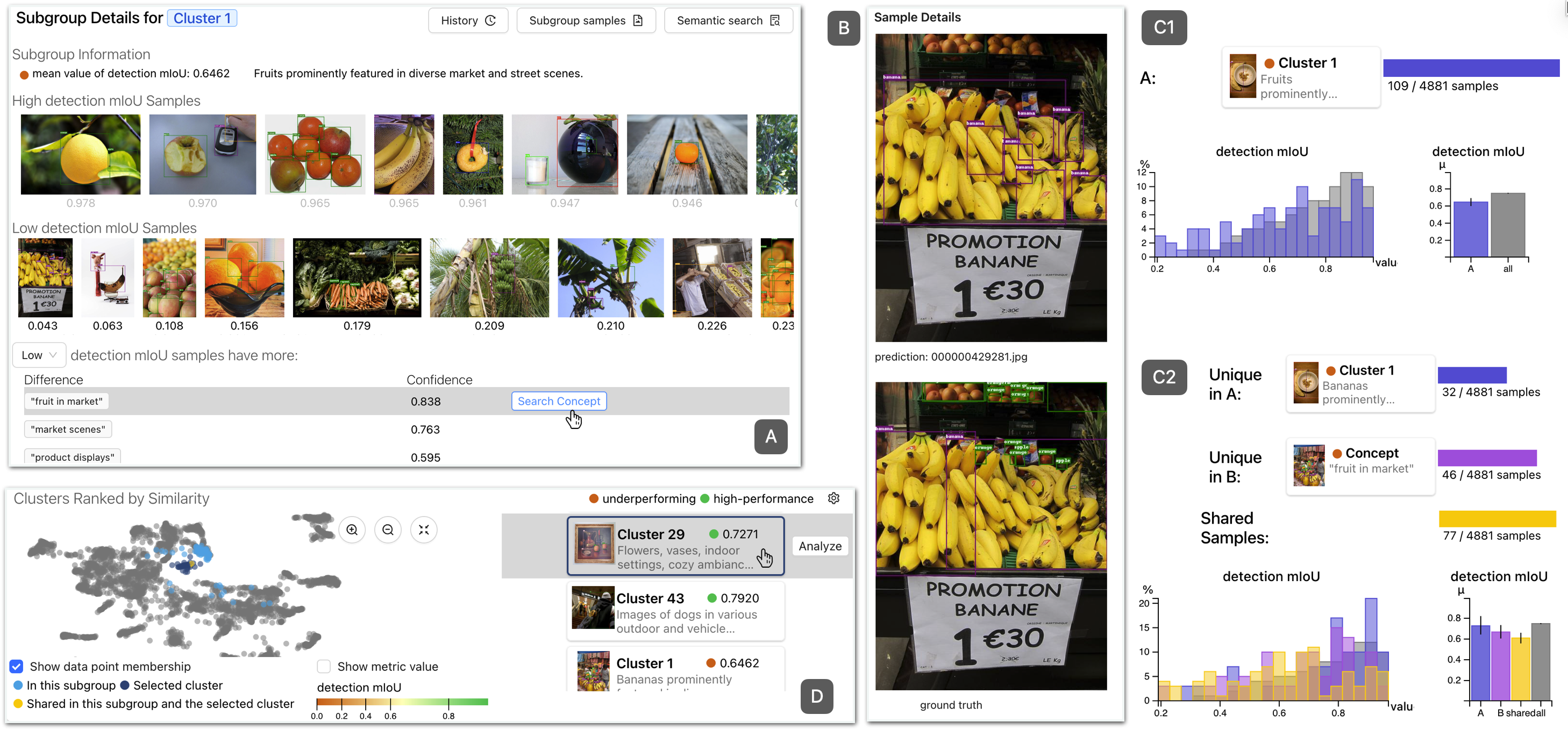}
  \caption{Using {\systemname} to analyze an object detection model. After comparing Cluster 1 and the subgroup of samples with ``fruit in market'' as suggested by GPT-4, E5 confirmed the labeling issues of inconsistency when a group of similar objects appear in the image.
  }
  \label{fig:ui_object}
\end{figure*}

\textbf{Analyzing a cluster with inconsistent labels.} 
E5 aimed to evaluate the detection performance of the pre-trained model by generating 50 semantic clusters using K-Means and ranking them by \textit{detection mIoU}, which measures the model's ability to detect objects without considering classification performance. E5 focused on \textit{Cluster 1}, which had the lowest average \textit{detection mIoU} values, indicating it contained many incorrect detection areas. As noted in the auto-generated summary, most samples in this cluster featured fruits found in markets and streets. She confirmed this observation by reviewing all subgroup samples.
Next, E5 selected one of the lowest-performing samples to compare the bounding boxes in the predictions and the ground truth (Fig.~\ref{fig:ui_object}B). She discovered label inconsistencies with bananas and oranges. Specifically, both the model predictions and the ground truth sometimes focused on individual fruits, such as a single banana, while other times they encompassed the entire area of the bananas. E5 explained that this inconsistency may not pose a problem if the model detects all the bananas' area. However, it becomes an issue if the task requires instance-level accuracy, where each object must be detected individually. E5 observed that this same problem occurred in other images within the cluster, affecting apples and oranges as well.

\textbf{Exploring a concept that may contain the same problem.} 
E5 aimed to explore the problem more deeply. Based on the auto-generated visual differences (Fig.~\ref{fig:ui_object}A), E5 searched for the concept of \textit{``fruit in market''}. E5 anticipated finding more samples with a similar annotation issue, where some fruits in a market tile were annotated individually while others were grouped. To refine the search, she adjusted the similarity threshold to filter out grocery/market images featuring other fruits.
Next, E5 checked how well these images represented the samples in \textit{cluster 1} and whether they contributed to the low \textit{detection mIoU} in that cluster. She added cluster 1 to the subgroup analysis panel (Fig.~\ref{fig:ui_object}, C1), revealing a higher percentage of low-performance samples compared to the entire dataset, as shown in the histogram. The average \textit{detection mIoU} for cluster 1 was significantly lower than that of the whole dataset.

E5 then included the subgroup of the concept \textit{``fruit in market''} in the analysis panel, highlighting the shared samples (Fig.~\ref{fig:ui_object}, C2). This concept encompassed 77 out of 109 samples in cluster 1. Moreover, the shared samples (yellow) accounted for the majority of low \textit{detection mIoU} cases, as indicated in the histogram. After removing these shared samples, the unique samples in cluster 1 (blue) showed no significant difference in mean \textit{detection mIoU} compared to the entire dataset. However, the shared samples (yellow) had a significantly lower average \textit{detection mIoU} than the unique samples (blue). E5 concluded that the samples related to \textit{``fruit in market''} primarily contributed to the lower performance of cluster 1, largely due to inconsistent bounding box labels.

\textbf{Investigating other clusters} 
To explore potential labeling issues, E5 investigated the nearest neighbors around this concept (Fig.~\ref{fig:ui_object}D). E5 identified cluster 29, which contained many samples of flowers displayed in markets or parks, and noted some samples with inconsistent labeling.
E5 then sorted the cluster table by \textit{detection mIoU} in descending order and analyzed several high-performance clusters. Upon reviewing the samples, E5 observed that these clusters included many images of individual animals (e.g., bears, zebras) or modes of transportation (e.g., airplanes). This further confirmed E5's finding that the primary issues in this dataset and model were related to inconsistent labels for groups of identical objects.

\textbf{Variations in participant analysis strategies.} 
E6 recognized the same labeling issues as E5 but was more interested in food samples. E6 explored various food categories, searching for terms like ``mixed food'', ``raw fruit'', and ``small food'' to analyze different types of food.

\section{Study Results Analysis \& Discussion}
\label{sec:results}
This section analyzes exploration patterns and expert feedback from our case studies, followed by reflections on lessons learned throughout the process. We hope these insights will inform the design of future visual analytics (VA) systems that leverage large foundation models.

\subsection{Study Results Analysis} 
\textbf{Exploration patterns.} All experts completed the high-level task of identifying where and why the model made errors using {\systemname}. However, their exploration patterns varied widely.
E1 and E2 used the \textit{Semantic Distribution Panel} more often, likely due to the smaller GSO validation set and clearer low-performance clusters. 5 out of 6 participants began by checking clusters in the \textit{Semantic Subgroups} table to get an overview of worst cases, then alternated between cluster exploration and concept search. E5, however, focused on one cluster before exploring neighboring subgroups. Due to time constraints, only E2 and E6 revisited clusters using \textit{History}.

\textbf{Complementary workflow integration} All experts found {\systemname} valuable for error analysis, especially for semantic search and hypothesis validation, \revise{which was hard with existing tools they were using}. For example, E6 easily identified “crab stick” outliers, previously hard to detect due to missing labels, while E4 spotted specific hairstyles that were challenging with other tools. E3 praised {\systemname} for enabling ``quick and dirty checks''.

Five participants appreciated the \textit{low performance vs. high performance} comparison in the \textit{Semantic Details} panel, which aided hypothesis generation. Auto-generated summaries and issue candidates helped E1 and E3 gain new insights from familiar datasets.
Three participants, who had focused on numeric metrics, appreciated the \textit{subgroup-level visualizations}. E2 noted that “semantic search alone was sufficient to quickly validate hypotheses” but used charts in the subgroup analysis panel for “more concrete evidence.”
E1 and E2 also emphasized the value of projection and semantic clustering, pointing out the lack of tools for systematic exploration.

\textbf{Suggestions for improvement.} Two participants requested more control over prompts, while three suggested a larger projection view in the \textit{Subgroup Analysis} page.

\subsection{\revise{Lessons Learned: Foundation Models in VA}}
We outline the benefits of using foundation models for visual analytics and address the associated risks.

\textbf{The role of system-generated summaries.}
Foundation models now enable hypothesis generation through system-generated subgroup summaries and candidate issues, complementing or, in some cases, replacing manual data analysis. In our study, five out of six participants generated hypotheses from these summaries. However, E3 preferred inspecting samples directly for more accurate insights, despite finding the summaries helpful.

\textbf{Awareness of uncertainty, ambiguity, and bias.}
Foundation models can occasionally produce vague outputs (e.g., ``people in various clothing''), which may hinder hypothesis formation. Bias in model embeddings is another concern in both VA and AI communities~\cite{kaddour2023challenges,slyman2023vlslice}. For instance, searching for ``makeup'' in the CelebA dataset mostly retrieved female images. Although participants recognized this bias, it did not significantly impact their work. E4 noted that ``this bias reflects the data'', while E2 acknowledged LLM uncertainty but still found foundation models useful for querying samples and identifying potential issues. All participants expressed confidence using \systemname{} for error analysis, as it allowed for adjusting search thresholds and manually refining results to minimize bias.

\textbf{Semantic analysis with minimal annotations.}
Participants emphasized the need for semantic analysis to understand model behavior and explain errors related to specific concepts. E2 noted the difficulty of analyzing models without rich annotations, relying instead on custom metrics that don’t fully capture error patterns. This highlights the value of foundation models in generating semantically meaningful data, offering deeper insights beyond traditional metrics, especially in datasets with limited labels.

\subsection{Lessons Learned: VA for Insight Validation}
Through user feedback and case studies, we confirmed the critical role of visual analytics in model debugging, with the following key insights:

\textbf{Visual comparison facilitates insights.}
Users rely on comparison to understand model errors. In the case study, all participants compared model predictions to ground truth, low-performance to high-performance samples, and subgroup metrics to the overall dataset. They also compared samples and metrics across subgroups and search results, highlighting the importance of intuitive comparison tools to support analysis.

\textbf{Visual confirmation for hypothesis testing.} 
While hypothesis testing is a well-discussed topic in visualization\cite{suh2022grammar}, there's limited support for model debugging. 
In the existing workflow of model debugging, E1-E4 just report the numeric values of key metrics over low-performance subgroups. E5 and E6 used basic visualization to support their analysis (e.g., bar chart of metrics over different classes). 
Our \textit{Subgroup Analysis} panel was seen as a valuable addition, making users more confident in their analysis by showing metric distribution differences. E3 and E6 also noted its usefulness for team communication.

\textbf{Visualizing samples aids embedding interpretation.}
As highlighted in a recent survey on model embedding usage~\cite{huang2023va}, there are notable challenges in employing model embeddings within a visual analytics (VA) system. One key issue is the bias inherent in the model embeddings. As discussed in the previous subsection, visualizing derived results allows users to identify and mitigate this bias. Another challenge arises from the mismatch between the high dimensionality of original embeddings and their 2D or 3D visual projections. E6 noted their awareness of the distortions introduced by dimensionality reduction but appreciated the ability to visually examine sample distributions. This visualization aids in assessing whether clusters and search-based subgroups are genuinely semantically similar. Additionally, being able to view actual outlier samples helps users understand how these data points differ from others.

\subsection{Reflection of Practical and Societal Impact}
As large models become increasingly integrated into workflows, it’s crucial to address their practical and societal impact, particularly around privacy, fairness, diversity. \revise{In the following sections, we elaborate how we protect the aforementioned values in the paper.}

\textbf{Privacy and Personally Identifiable Information (PII) Data Protection.} We address these concerns by sending queries to external LLMs (e.g. GPT-4) using captions derived locally from images, instead of transmitting the original images. This reduces the risk of exposing sensitive visual data. Additionally, we included in our system an image blur feature, allowing users to mask PII before sharing data with collaborators, ensuring compliance with privacy standards. These measures enable secure use of LLMs without compromising sensitive information.

\textbf{Fairness and Diversity Check.}
Ensuring fairness across subpopulations is another key feature of {\systemname}. Users can easily retrieve samples based on demographic attributes like gender, age, and skin tone, as described in Sec.~\ref{sec:case_face}. This functionality helps evaluate model performance across diverse groups, ensuring that models do not unfairly favor or harm specific populations. It also encourages developers to diversify their training data, promoting inclusivity and preventing underrepresentation.

\revise{
\textbf{Flexibility of Model Selection.} The workflow we propose incorporates a highly flexible model selection mechanism, allowing seamless integration of state-of-the-art models tailored to user preferences and domain-specific needs. The implemented system is designed to support modularity, allowing users to replace existing models with alternatives that better align with their goals, whether it involves improved performance, specialized functionality, or adherence to data compliance and legal requirements. This adaptability ensures the workflow remains applicable across various domains, empowering users to leverage the latest advancements in model development without being constrained by the original system's default configurations.
}
\section{Limitations and Future Work}
\label{sec:limit}
In the current system, there remain a couple of limitations that we seek as opportunities for future iteration to improve on. The first limitation comes from the CLIP-based semantic search. We noticed that currently, CLIP model cannot support search with negation (e.g., ``not wearing hat'') or numbers (e.g., ``two bananas'') well. We would like to explore other multi-modality models in the future.
% In addition, by only running case studies with six model developers (expert users), we may be overgeneralizing our results. As such, evaluating {\systemname}'s effectiveness in diverse application domains and with varying user expertise levels would provide valuable insights for refinement and optimization. 
\revise{
For the evaluation of \systemname, we conducted in-depth interviews with only six experts, as this method is well-suited for the usage scenario of CVML model analysis which requires specialized expertise of model development. These interviews allowed us to gather valuable observations and insights about the system’s utility and limitations. While this approach provided rich qualitative insights, a future quantitative study with more people will be necessary to validate the findings across diverse application domains and user expertise levels.
}

Based on feedback from domain experts, we plan to extend {\systemname} to accommodate a broader range of computer vision tasks and data that would enhance its applicability and utility. Additionally, enabling customized prompts for subgroup summarization would empower users to tailor the analysis to specific research questions or application domains, thereby increasing flexibility and adaptability.

\section{Conclusion } 
\label{sec:conclude}
In this work, we introduce \systemname, a novel visual analytics workflow for subgroup-based semantic error analysis of CVML models. Through formative interviews with seven machine learning engineers and hierarchical task analysis, we identified two key challenges in semantic error analysis and outlined corresponding design goals. Based on these, we combine the use of large foundation models with visual analytics techniques to help analysts gain deeper insights into error patterns and validate hypotheses, and illustrate how such workflow can be integrated in a practical system. Our case studies and expert interviews \revise{highlight the complementary role of \systemname{} in existing error analysis workflows. From our discussions, we distilled several key lessons, including the usage of large foundation models within visual analytics systems and the impact of visual analytics for insight validation.} 

% demonstrate the effectiveness of \systemname{} in uncovering error trends and supporting informed decision-making for model improvement.

% references
\bibliographystyle{ACM-Reference-Format}
\bibliography{ref}

\end{document}